\documentclass[sigconf]{acmart} %publication (i.e., after conditional acceptance) in the DL

\usepackage[utf8]{inputenc}
\usepackage[most,breakable]{tcolorbox}
\usepackage[inline]{enumitem}
\usepackage{multirow}
\usepackage{bbding}
\usepackage{makecell}
\usepackage{tabularx}
\usepackage{comment}

\AtBeginDocument{%
  \providecommand\BibTeX{{%
    \normalfont B\kern-0.5em{\scshape i\kern-0.25em b}\kern-0.8em\TeX}}}

\setcopyright{acmcopyright}
\copyrightyear{2018}
\acmYear{2018}
\acmDOI{XXXXXXX.XXXXXXX}

\acmConference[Conference 'XX]{Make sure to enter the correct
  conference title from your rights confirmation emai}{June 03--05,
  2018}{XX, XX}
\acmBooktitle{Woodstock '18: ACM Symposium on Neural Gaze Detection,
 June 03--05, 2018, Woodstock, NY} 
\acmPrice{15.00}
\acmISBN{978-1-4503-XXXX-X/18/06}

\begin{document}
\makeatletter
\renewcommand{\fnum@table}{\tablename~\thetable}
\makeatother
%\title{StoryDiffusion: Exploring How Designers Co-Create Storyboards With a Generative-AI System}%StoryDiffusion: Designers' Journey Creating Storyboards with Generative AI}
% \title{StoryDiffusion: How Designers Co-Create Storyboards With a Generative-AI System}
% \title{StoryDiffusion: How to Support UX Storyboard Creation With Generative-AI}
\title{StoryDiffusion: How to Support UX Storyboarding With Generative-AI}

\author{Zhaohui Liang}
\affiliation{%
  \institution{University of Chinese Academy of Sciences}
  \institution{Computer Network Information Center, Chinese Academy of Sciences}
  %\streetaddress{P.O. Box 1212}
  \city{Beijing}
  \country{China}
}
\email{zhliang@cnic.cn}

\authornote{Both authors contributed equally to this research.}
\author{Xiaoyu Zhang}
\authornotemark[1]
\affiliation{
  \institution{Beijing Institute of Technology}
  %\streetaddress{P.O. Box 1212}
  \city{Beijing}
  \country{China}
}
\email{3120211891@bit.edu.cn}

\author{Kevin Ma}
\affiliation{%
  \institution{University of California, Berkeley}
  \city{Berkeley}
  \state{California}
  \country{USA}
}
\email{kevinma1515@berkeley.edu}

\author{Zhao Liu}
\affiliation{%
  \institution{Beijing Institute of Technology}
  \city{Beijing}
  \country{China}
}
\email{zhaoyoung6@outlook.com}

\author{Xipei Ren}
\affiliation{%
  \institution{Beijing Institute of Technology}
  \city{Beijing}
  \country{China}
}
\email{x.ren@bit.edu.cn}

\author{Kosa Goucher-Lambert}
\affiliation{%
  \institution{University of California, Berkeley}
  \city{Berkeley}
  \state{California}
  \country{USA}
}
\email{kosa@berkeley.edu}

\author{Can Liu}
\authornote{Corresponding Author.}
\affiliation{%
  \institution{School of Creative Media, City University of Hong Kong}
  %\streetaddress{8600 Datapoint Drive}
  \city{Hong Kong}
  %\state{Texas}
  \country{China}
  %\postcode{78229}}
}
\email{canliu@cityu.edu.hk}

\renewcommand{\shortauthors}{Zhaohui and Xiaoyu, et al.}

\begin{abstract}
Storyboarding is an established method for designing user experiences. Generative AI can support this process by helping designers quickly create visual narratives. However, existing tools only focus on accurate text-to-image generation. Currently, it is not clear how to effectively support the entire creative process of storyboarding and how to develop AI-powered tools to support designers' individual workflows. In this work, we iteratively developed and implemented StoryDiffusion, a system that integrates text-to-text and text-to-image models, to support the generation of narratives and images in a single pipeline. With a user study, we observed 12 UX designers using the system for both concept ideation and illustration tasks. Our findings identified AI-directed vs. user-directed creative strategies in both tasks and revealed the importance of supporting the interchange between narrative iteration and image generation. We also found effects of the design tasks on their strategies and preferences, providing insights for future development.
\end{abstract}

%




\begin{teaserfigure}
  \centering
  \includegraphics[width=0.85\linewidth]{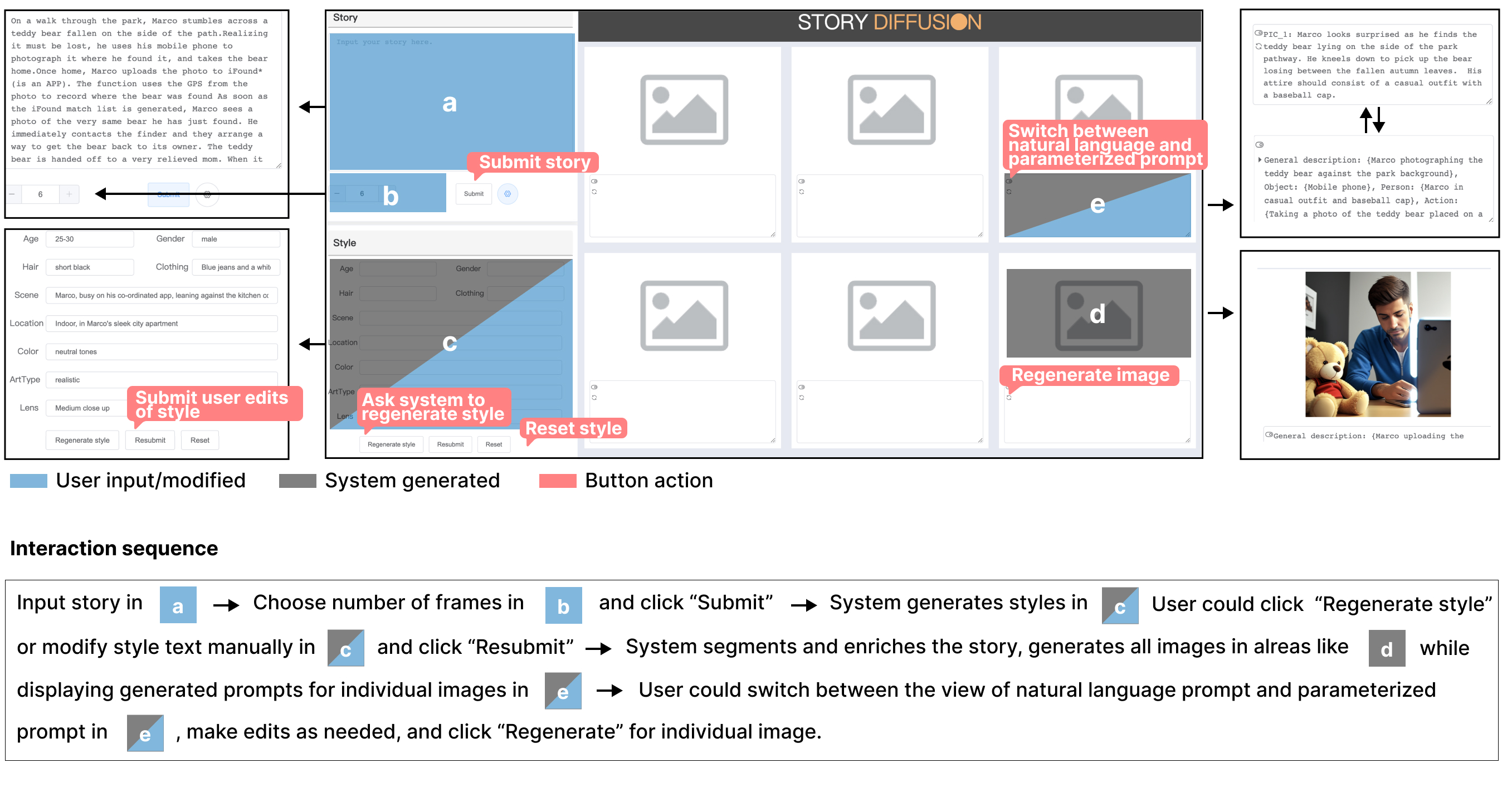}
  \caption{The figure displays the workflow for StoryDiffusion ordered in steps a, b, c, d and e. In our workflow, we utilized GPT-4 and a stable diffusion model to allow users to input brief ideas or complete story descriptions for the system to automatically generate an entire storyboard. Designers can then adjust and refine each frame (image) using natural language and prompt inputs.}
  \Description{}
  \label{WORKFLOW}
\end{teaserfigure}
% \begin{figure}
%   %\centering
%   \includegraphics[width=\linewidth]{Fig/HCIX-WORKFLOW-05.jpg}
%   \caption{The workflow for using StoryDiffusion.}
%   \Description{}
%   \label{WORKFLOW}
% \end{figure}
\received{20 February 2007}
\received[revised]{12 March 2009}
\received[accepted]{5 June 2009}

\maketitle

\section{Introduction}

During the User Experience (UX) design process, storyboards serve as an important tool for designers to visualize the user journey and empathize with users. The visual narratives can enhance UX designers' ability to ideate conceptual design solutions and communicate their concepts effectively to stakeholders \cite{mohd2014review, truong2006storyboarding, van2006value}. In this context, the recent advances in generative artificial intelligence (GAI) present an opportunity to improve the storyboarding process \cite{epstein2023art}. Thus, designers can leverage GAI to co-create visual narratives that are highly detailed, dynamic, and adaptable \cite{antony2023id, shi2020emog, han2023design}. As a result, researchers have explored using GAI to aid in performing this task \cite{li2019storygan, liu2022design, maharana2022storydall}, and their findings show that GAI is capable of accelerating the design task of creating storyboards. 

Existing systems for AI-assisted storytelling either focus on script writing or frame-by-frame text-to-image generation \cite{mirowski2023co, rao2023dynamic}. However, storyboarding is a task that involves both narrative and visual creation. It has been less explored how to support designers through the entire creative process involving both aspects. Furthermore, prior research on AI-generated storyboards mainly concentrated on building models and systems, with little exploration into how designers would apply their system to various storyboard-related design activities \cite{li2019storygan, maharana2022storydall, shi2020emog}. To address these gaps, this work sets out to explore how GenAI models can be used to support the entire storyboarding activity, as well as how they can be integrated into designers' diverse workflows for different tasks. 

Building on existing work, we designed and developed StoryDiffusion, a system that harnesses the recent advancements in text-to-text and text-to-image GAI models by integrating them within one system through a step-wise approach \cite{chakrabarty2023spy, jeong2023zero}. It allows users to input an overarching narrative, either detailed or abstract, from which a sequence of storyboard frames is generated. Designers can then refine each frame by adjusting multiple prompt inputs, striking a balance between automation and user control. 
It is known that storyboards should be composed of coherent sequences of images representative of the user narrative \cite{truong2006storyboarding}. Towards this goal, we used a novel system that involved prompting GPT-4 \cite{openai2023gpt4} to segment the designer-inputted storylines and translate them into parameterized prompts that work better as input for text-to-image models. These prompts are then inputted into a text-to-image model to generate consistent and temporally coherent frames.

% e recognize that a designer's use case and workflow for a storyboard can vary, so we set out to explore how GenAI models can be integrated into their workflows for various design tasks.
Prior work has shown that storyboards can be used by designers to illustrate a user's journey to communicate their product idea to key stakeholders (concept demonstration), or they can be used by designers as a tool for ideation (concept ideation) \cite{ulrich2008product, truong2006storyboarding}. To gain a comprehensive understanding of how UX designers could integrate StoryDiffusion into their workflows, %we conducted a user study with designers pocessing different levels of expertise.Thus,
we conducted a lab study with design-majored college students experienced in UX storyboarding, to observe their experience using StoryDiffusion for both types of design tasks: \textit{concept ideation} and \textit{concept illustration}. In concept ideation, designers interacted with StoryDiffusion to brainstorm potential ideas for an undefined product, while in concept illustration, designers created a storyboard to demonstrate a well-defined product concept. This exploration allowed us to identify and articulate key strategies that designers employed while utilizing StoryDiffusion across different design tasks.
% Our user study aimed to gain a comprehensive understanding of how designers integrated StoryDiffusion into their workflow in both design tasks. 
Our analysis aimed to identify the needs and strategies that emerged when designers interacted with StoryDiffusion, which we anticipate will inform the future evolution of GAI-storyboard systems.

As a result, our paper's main contributions are as follows:

\begin{itemize}
    \item We built a system that integrates an existing pre-trained text-to-text model (GPT-4) and a text-to-image model (Stable Diffusion \cite{automatic1111_2023, StableDiffusion2024}) to transform a designer's textual narrative, which can be in any level of detail, into a series of six images, laying a groundwork for future GAI-enhanced storyboard tools.
    \item We observed how design students used StoryDiffusion in two distinct storyboarding tasks: concept ideation and concept illustration. This investigation led to the identification and discussion of key strategies adopted by designers during their interaction with StoryDiffusion.
    \item We discussed the implications of our findings by shedding light on how storyboards that utilize GAI can influence the designer's process of creating storyboards. In addition, we discussed how different design tasks affected the use of StoryDiffusion. These insights aim to inform the development of future systems alike. %by identifying and studying specific designer needs and strategies.
\end{itemize}

The structure of our paper is as follows. First, we explored the essential elements for effective storyboards discussed in both the formative study and related works. %Particularly, we noted designers needed consistency across each frame (images) in a storyboard. 
In addition, we summarized the prior research aimed at resolving similar issues in storyboard generation \cite{hertz2022prompt, jeong2023zero, li2019storygan, liu2023intelligent, pan2022synthesizing}. We then discussed our system's architecture and the protocol for our user study. Following this, we presented the results from our user study and discussed its implications for future works.

\section{Related Work}
\label{related_works}

\subsection{GAI to support designers}
\label{creative_support}
The field of GAI, with innovations such as OpenAI's DALL-E and ChatGPT, has led to a surge in research exploring its potential to augment the creative processes of users. In one study, GAI has shown to be capable of serving as a reflective tool, enabling designers to gain deeper insights into their experiences \cite{canet2022dream}. This notion aligns with the perspective that designers can harness GAI to better understand their design problem, therefore enhancing their creative output. 

Previous studies have provided evidence that tools incorporating GAI can bolster design quality \cite{verheijden2023collaborative, zhong2024ai, vaithilingam2024imagining}. Moreover, there are increasing calls for a symbiotic, co-creative relationship between designers and AI \cite{hwang2022too, vinchon2023artificial, chung2021intersection}. This body of work investigates the collaborative dynamic whereby AI and designers utilize their respective strengths to function as a cohesive unit in the creative process. This idea requires the recognition of AI's proficiency in tasks that are unfeasible for humans, such as providing intricate artwork and stories within a couple of seconds, and the ability to exploit AI's capabilities for those tasks such that humans can assume the role of managing, evaluating, and curating generated ideas into a single idea \cite{inie2023designing, brade2023promptify}. 

In general, the literature indicates that AI excels at generating a plethora of personalized potential outputs for humans to utilize \cite{vinchon2023artificial}. Thus, this personalization of AI-generated artifacts allows humans to engage in higher-order decision-making, relegating the AI to the role of executing the more routine and technical aspects of idea generation \cite{vinchon2023artificial}. Drawing parallels to the field of design, these findings suggest that designers can view AI-generated outputs as a source of inspiration, aiding them during the ideation phase \cite{goucher2019crowdsourcing, kwon2022exploring}. This approach, which can also be termed as ``design inspiration search'', positions AI as facilitators in divergent thinking tasks, where designers are tasked with exploring and brainstorming a breadth of ideas \cite{lu2022bridging}. On the other hand, AI can also assist designers in convergent thinking tasks, which are characterized by the need for precise decision-making \cite{hwang2022too}. In this paper, we studied StoryDiffusion for both concept ideation and concept illustration, which will expand on prior work by enabling us to study how designers incorporate GAI storyboarding tools into their workflow for similar design tasks.

\subsection{GAI in storyboarding}
\label{relatedworks:genAI}
Prior research has resulted in a variety of storyboard tools tailored specifically to aid the design process. For instance, Emog is a GAI system designed to help designers incorporate emotional states into a storyboard, which addresses the limitations of previously existing generative storyboard systems that lack that capability \cite{shi2020emog}. Similarly, tools like Storycanvas and Storeoboard were created to make storyboards more accessible to designers with limited artistic or technical skills \cite{henrikson2016storeoboard, skorupski2010novice}. 

GAI has emerged as a powerful tool to automate the process of creating a storyboard. This evolution is largely attributed to advances in the development of AI models, such as generative adversarial networks, diffusion models, and transformers \cite{goodfellow2020generative, rombach2022high, vaswaniattention2017}. Consequently, a variety of specialized GAI models and systems have been developed specifically for generating storyboards, such as StoryDALL-E \cite{maharana2022storydall}, ARLDM \cite{pan2022synthesizing}, Storia \footnote{https://app.storia.ai/}, Krock \footnote{https://krock.io/storyboard-ai/}, Boords \footnote{https://boords.com/ai-storyboard-generator} and storyboarder \footnote{https://storyboarder.ai/} (as shown in Table \ref{AISTORYBOARDINGTOOLS}). 
Notably, `Storia' and `StoryDALL-E' are systems that allow users to input and modify narratives on a frame-by-frame basis \cite{maharana2022storydall}. In addition, `ARLDM' is a model architecture developed to generate several temporally coherent image sequences from a single narrative input \cite{pan2022synthesizing}. Furthermore, `Boords' represents a product that provides two separate tools one for writing scripts and the other for generating storyboards. The script-writing tool integrates pre-trained LLMs with a pre-trained text-to-image model to transform a narrative into six sequential scripts. The storyboarding tool takes segmented scripts as input and generates individual storyboard frames for each. Thus, there exists a substantial body of prior research that demonstrates the capability of GAI to automate the creation of storyboards.

\begin{table*}[]
\caption{Comparison of existing GAI storyboarding tools.}
\renewcommand\arraystretch{1}
\scalebox{0.69}
{\begin{tabular}{|cc|cc|cccc|c|}
\hline
\multicolumn{2}{|c|}{Type} &
  \multicolumn{2}{c|}{Research} &
  \multicolumn{4}{c|}{Commercial Tools} &
  \multirow{2}{*}{Our tool} \\ \cline{1-8}
\multicolumn{2}{|c|}{Tool Name} &
  \multicolumn{1}{c|}{StoryDALL-E \cite{maharana2022storydall}} &
  ARLDM \cite{pan2022synthesizing} &
  \multicolumn{1}{c|}{Storia} &
  \multicolumn{1}{c|}{Krock} &
  \multicolumn{1}{c|}{Boords} &
  storyboarder &
   \\ \hline
\multicolumn{1}{|c|}{\multirow{2}{*}{Model}} &
  \begin{tabular}[c]{@{}c@{}}text-to-text \\ model\end{tabular} &
  \multicolumn{1}{c|}{\XSolid} &
  \XSolid &
  \multicolumn{1}{c|}{\XSolid} &
  \multicolumn{1}{c|}{\XSolid} &
  \multicolumn{1}{c|}{\Checkmark} &
  \Checkmark &
  \Checkmark \\ \cline{2-9} 
\multicolumn{1}{|c|}{} &
  \begin{tabular}[c]{@{}c@{}}text-to-image \\ model\end{tabular} &
  \multicolumn{1}{c|}{\Checkmark} &
  \Checkmark &
  \multicolumn{1}{c|}{\Checkmark} &
  \multicolumn{1}{c|}{\Checkmark} &
  \multicolumn{1}{c|}{\Checkmark} &
  \Checkmark &
  \Checkmark \\ \hline
\multicolumn{1}{|c|}{\multirow{5}{*}{Interaction}} &
  Edit story &
  \multicolumn{1}{c|}{-} &
  - &
  \multicolumn{1}{c|}{\Checkmark} &
  \multicolumn{1}{c|}{\Checkmark} &
  \multicolumn{1}{c|}{\Checkmark} &
  \Checkmark &
  \Checkmark \\ \cline{2-9} 
\multicolumn{1}{|c|}{} &
  \begin{tabular}[c]{@{}c@{}}Choose number \\ of frames\end{tabular} &
  \multicolumn{1}{c|}{-} &
  - &
  \multicolumn{1}{c|}{\Checkmark} &
  \multicolumn{1}{c|}{\Checkmark} &
  \multicolumn{1}{c|}{\Checkmark} &
  \XSolid &
  \Checkmark \\ \cline{2-9} 
\multicolumn{1}{|c|}{} &
  Edit style &
  \multicolumn{1}{c|}{-} &
  - &
  \multicolumn{1}{c|}{\Checkmark} &
  \multicolumn{1}{c|}{\Checkmark} &
  \multicolumn{1}{c|}{\Checkmark} &
  \XSolid &
  \Checkmark \\ \cline{2-9} 
\multicolumn{1}{|c|}{} &
  \begin{tabular}[c]{@{}c@{}}Edit description \\ for single image\end{tabular} &
  \multicolumn{1}{c|}{-} &
  - &
  \multicolumn{1}{c|}{\XSolid} &
  \multicolumn{1}{c|}{\XSolid} &
  \multicolumn{1}{c|}{\XSolid} &
  \Checkmark &
  \Checkmark \\ \cline{2-9} 
\multicolumn{1}{|c|}{} &
  \begin{tabular}[c]{@{}c@{}}Edit prompt \\ for single image\end{tabular} &
  \multicolumn{1}{c|}{-} &
  - &
  \multicolumn{1}{c|}{\XSolid} &
  \multicolumn{1}{c|}{\XSolid} &
  \multicolumn{1}{c|}{\Checkmark} &
  \XSolid &
  \Checkmark \\ \hline
\multicolumn{1}{|c|}{\multirow{2}{*}{\begin{tabular}[c]{@{}c@{}}Prompt \\ Engineering\end{tabular}}} &
  Style &
  \multicolumn{1}{c|}{-} &
  - &
  \multicolumn{1}{c|}{\begin{tabular}[c]{@{}c@{}} 10 movie styles\end{tabular}} &
  \multicolumn{1}{c|}{\begin{tabular}[c]{@{}c@{}}pencil, \\ watercolour, \\ 3D render, \\ photo realistic\end{tabular}} &
  \multicolumn{1}{c|}{\begin{tabular}[c]{@{}c@{}}21 \\ art \\ styles\end{tabular}} &
  low-fidelity &
  \begin{tabular}[c]{@{}c@{}}human-AI co-creation \\ style within \\ a template \\ contain 9 parameters\end{tabular} \\ \cline{2-9} 
\multicolumn{1}{|c|}{} &
  Parameters &
  \multicolumn{1}{c|}{-} &
  - &
  \multicolumn{1}{c|}{\XSolid} &
  \multicolumn{1}{c|}{\begin{tabular}[c]{@{}c@{}}action, \\ narrator, \\ camera\end{tabular}} &
  \multicolumn{1}{c|}{\begin{tabular}[c]{@{}c@{}}character,\\ action, \\ narrator, \\ camera movement\end{tabular}} &
  \begin{tabular}[c]{@{}c@{}}shot size,\\ perspective, \\ movement, \\ equipment, \\ focal length, \\ filter\end{tabular} &
  \begin{tabular}[c]{@{}c@{}}9 parameters \\ on general level \\ and 8 parameters \\ on image level\end{tabular} \\ \hline
\end{tabular}}
\label{AISTORYBOARDINGTOOLS}
\end{table*}

Our system shares similarities with these works but introduces a novel approach by supporting the narrative development and image generation in one tool. As seen in Table \ref{AISTORYBOARDINGTOOLS}, we also provide more access points for editing in the whole process as well as generating more parameters tailored for UX design storyboards through our prompt engineering method. With StoryDiffusion, designers can input a narrative, either with a short summary or with elaborated details, and the system generates an entire storyboard. If the story lacks detail, the system leverages a task-oriented prompting method to fill in the details. Designers then have the flexibility to revise the story, or revise the description / parameterized prompt for each frame. % allowing for targeted modifications of each respective image. However, we still had to address the challenge of prompt engineering across multiple models to align the generated storyboard with the designer's narrative input. To address this, 

%In our system, we leveraged GPT-4 \cite{openai2023gpt4} to transform the designer's textual narrative into six prompts (can be more or less) that are inputted into StoryDiffusion, which then generates images that represent the designer's narrative. We note that an additional advantage of our system as opposed to prior work is the accessibility of the underlying LLM and text-to-image models used to build the system, which we anticipate will make our system more accessible for future practitioners and researchers seeking to develop GAI storyboard systems.

\section{Formative Study - how can GAI support storyboarding}\label{formative_study}

% \subsection{Factors for Effective Storyboards}
% \label{factors_effective}
In product design, storyboards can serve as a tool for visualizing narratives and communicating design ideas amongst stakeholders \cite{ulrich2008product, van2006value}. They can also enable designers to convey ideas effectively and foster empathy with the users by illustrating different scenarios of user-product interaction. Recognizing the dual utility of storyboards, our research investigates StoryDiffusion's application in both ideation and illustration contexts \cite{haesen2010draw}.%, exploring the designer's journey in each scenario.

Previous work summarized the essential elements that constitute an effective storyboard: detail and context, people and emotions, sequential frames, and depiction of temporal progression \cite{truong2006storyboarding, haesen2010draw, peng2017storytelling}. Our system aims to incorporate these elements to enhance the quality and usability of our AI-generated storyboards. Although this knowledge helps us design the automation of generating higher quality storyboards, what remains unknown is what support UX designers need during their creative process. Thus we conducted a formative study to identify additional needs and requirements for AI-generated storyboards.

% In Section \ref{factors_effective}, we pulled from prior literature some factors that lead to an effective storyboard, including detailed context, human emotions, sequential framing, and clear temporal progression \cite{truong2006storyboarding}. %Even though prior research has listed out some criteria that are required in storyboards, we argue that it is important to reassess and update our understanding of these needs given the rapid advancements in LLMs and text-to-image models. 
% Although this knowledge helps us design the automation of generating higher quality storyboards, what remains unknown is what support UX designers need during their creative process. Thus, we conducted a formative study to gain early insights on this to inform our design. % This formative study is driven by the necessity to refine, validate, and expand upon the existing framework of user needs to inform the design of StoryDiffusion and our user study.

\subsection{Study design}

As we aim to support both the narrative and visual creation of storyboarding, our design process started by exploring how to combine the use of a Large Language Model and a text-to-image generation model to aid the entire storyboarding process.
We recruited six UX design students (1 male, 5 females), aged 23 to 28 (M=25.17, SE = 0.95), to participate in the study. The study consisted of two parts: an interview and a task. In the interview, the participants were asked about their experiences in creating storyboards. Questions such as ``What are the objectives and processes involved in your usual storyboard creation?'' and ``What aspects do you prioritize when creating a storyboard?'' were asked. They were also asked to bring and show a recent storyboard they had created for class or work and the role of GAI in the storyboard creation process. The task required the participants to recreate their previously made storyboards using ChatGPT and Stable Diffusion while articulating their requirements and expectations. 
% a series of studies, each consisting of three 30-minute sessions. Each participant was informed that they were allowed to withdraw at any stage. Participants, referred to as FP01, FP02,.., and FP06, were selected to address three components. The first and second components consisted of interviews, while the third phase involved performing a task. They are as follows. (1) Background: Participants described their experiences in creating storyboards. Questions such as "What are the objectives and processes involved in your usual storyboard creation?" and "What aspects do you prioritize when creating a storyboard?" were asked. (2) Interview: Discussion on the content of a recent storyboard they had created for class or work and the role of GAI in the storyboard creation process. Questions like "Could you present or describe a storyboard you recently created?" and "How do you think computer-aided technologies could support your storyboard creation?" were asked. (3) Task and feedback: Participants recreated their previously made storyboards using GAI tools while articulating their requirements and expectations. 

\begin{figure}
  %\centering
  \includegraphics[width=\linewidth]{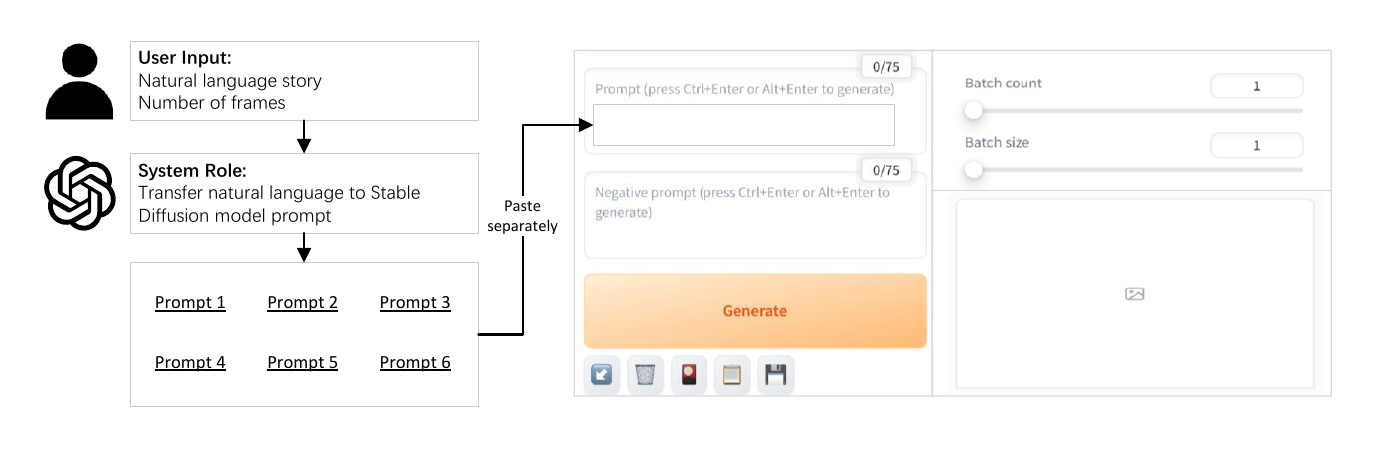}
  \caption{Workflow of the experiment used in the formative study. Users inputted their stories into ChatGPT to generate prompts, and then copied the prompts into a Stable Diffusion model to obtain images. System prompts for ChatGPT were given by us to define the role GPT-4 will play, which involved segmenting the story into several scenes based on the user's required number and generating corresponding prompts for the stable diffusion model. }
  \Description{}
  \label{formative_study_interface}
\end{figure}

% In the experimental component of our study (3), participants engaged in a process that simulated the generation of a storyboard provided that they already had a narrative of a user experience for a particular concept. Initially, 
In the task, participants were asked to articulate the narrative of their storyboards in the form of text, which were used as prompts to generate more prompts (based on the number of frames desired) through GPT-4. These prompts were subsequently inputted into a pre-trained text-to-image model, StableDiffusion \cite{automatic1111_2023, StableDiffusion2024}, to produce individual images as depicted in Figure \ref{formative_study_interface}. The participants were told to assemble these images into coherent sequence to create one storyboard.
During this process, participants were encouraged to repeat any steps to iteratively improve the storyboard. Following this, we solicited participants' feedback regarding their expectations versus the actual outcomes of the generated visuals and any concerns they had with the images produced. %Through this feedback, we gathered insights into participant's perceptions of the AI-generated storyboards and their thoughts on potential uses for automating natural language narratives into visual storyboards.

\subsection{Findings}
\begin{figure} 
  %\centering
  \includegraphics[width=\linewidth]{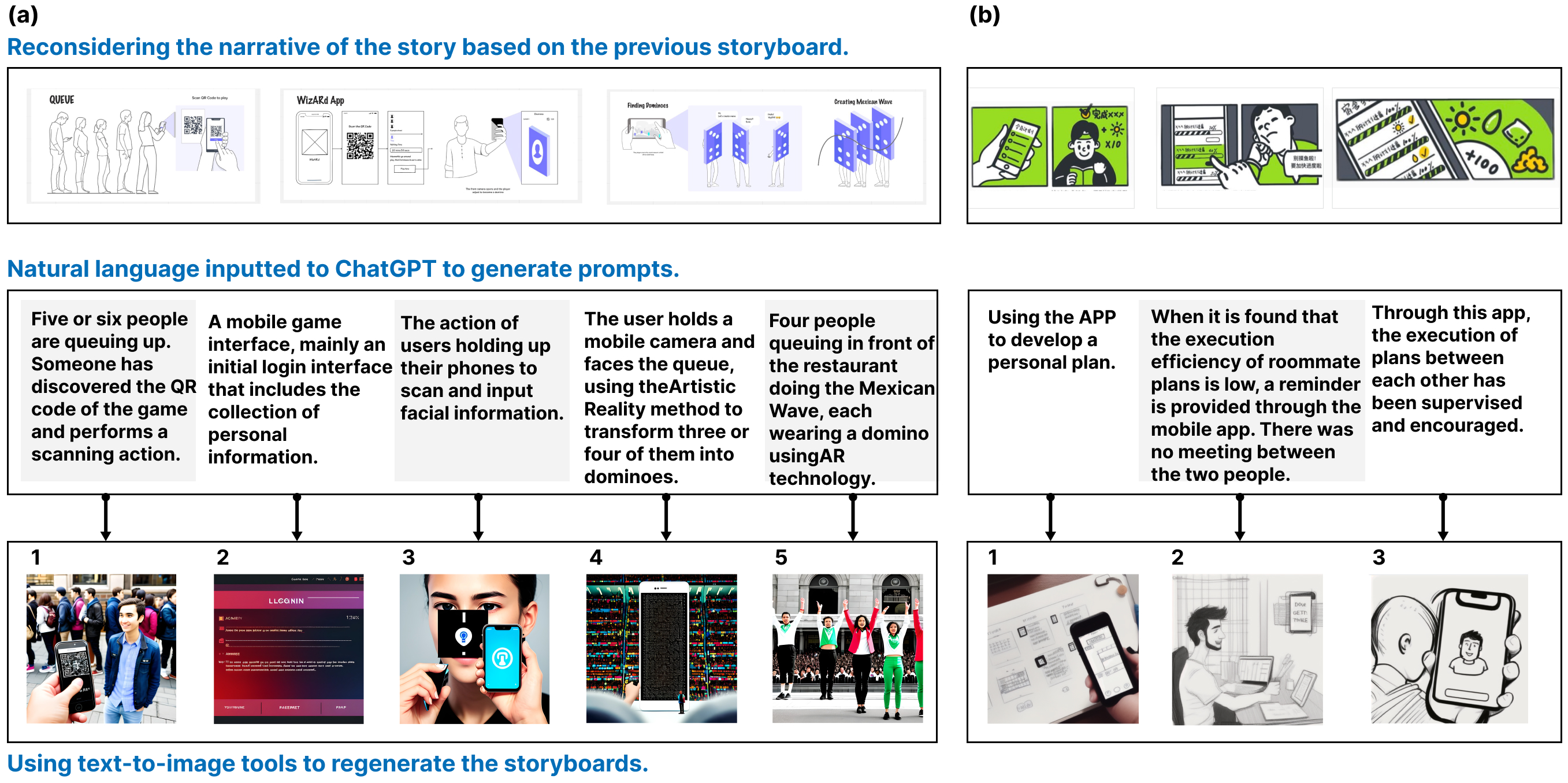}
  \caption{Storyboards regenerated by (a)FP02 and (b)FP05. Images in the first row were the storyboards drawn by the participants themselves, and images in the third row were the storyboards regenerated using text-to-image tools. The texts between two rows were the sentences used to generate prompts for each image.}
  \Description{}
  \label{INTERVIEW}
\end{figure}

Figure \ref{INTERVIEW} showed some examples of storyboards generated by the participants. In this study we identified usage cases where the designers found GAI to be useful in storyboard creation, gathered insights into their perceptions. We also gathered their thoughts on potential uses for automating natural language narratives into visual storyboards and noted some user needs given the current capabilities of existing pre-trained GAI models. These insights guided the development of StoryDiffusion and shaped the methodology of our user study. We present them in the following themes.

\subsubsection{Usage of GAI for UX storyboarding}
\paragraph{Design Ideation and Conceptual Illustration} Participants emphasized that GAI could be potentially useful for early stage design ideation and conceptual illustration of user experiences. For instance, Participant FP03 stated: \textit{``I drew two storyboards of my previous design, one illustrating the design problem, and the other the solution''}. Similarly, Participant FP05 stated that he could envision the storyboards being used for \textit{``explaining the design opportunities''} and to \textit{``explain my design''}. In both cases, the participants shared that GAI was useful for not only illustrating the solution but also exploring alternative design opportunities.

\paragraph{Leveraging AI for Creativity.} We observed that participants utilized GAI to help them spark creative ideas, and this was particularly valued by the participants. This included its application in generating visual effects, emotions, and character actions to aid ideations. One instance of this is when participant FP04 mentioned: \textit{``I am not good at drawing characters, but I would like to create a protagonist that is a cute little girl''}. This finding aligns with existing literature that advocates for the adoption of GAI as a collaborative tool for designers, supporting them in creative endeavors \cite{hwang2022too, vinchon2023artificial}.

\subsubsection{User Requirements}
  
\paragraph{Narrative Clarity and Visual Consistency.} When creating storyboards, participants underscored the importance of narrative clarity and visual consistency in their storyboards, echoing findings from prior research on the characteristics of effective storyboards\cite{truong2006storyboarding}. For instance, FP05 encountered difficulties in maintaining narrative clarity between images while creating storyboards. FP05 mentioned, "When drawing storyboards, I prefer to use multiple panels and include captions to explain pain points. I also emphasize the use of color to highlight key points for quick comprehension. Furthermore, I focus on portraying interactions between characters, objects, and the environment." Our proposed system, StoryDiffusion, addressed some of the specific challenges FP05 and other participants highlighted to enhance the storyboard creation process for designers.
    
\paragraph{Addressing the Human-AI Communication Gap.} Participants highlighted a significant discrepancy between their communicated intentions and the outputs produced by the GAI. For instance, FP06 mentioned, "I needed to describe things very precisely. I wish there could be a revision process. This would allow me to continue modifying my input after generating images." This suggests that certain key concepts envisioned by participants, such as ``queue'' and ``Artistic Reality'' in the second row of Figure \ref{INTERVIEW}, were not effectively captured by the GAI model even after several attempts. This discrepancy was evident when participants voiced challenges in achieving consistency in temporal and artistic styles. Consequently, participants frequently found themselves employing a variety of prompts to produce images that suit their needs, thereby increasing the complexity of interaction with the system.

The insights from this study, combined with prior work on effective storyboards, informed the development of StoryDiffusion. In the subsequent sections, we detailed the system's design and its alignment with the results from our formative study.

\section{Designing StoryDiffusion}
\label{methods_system}

In the formative study, we observed that users tend to compose stories with missing details but hope for a system-generated visual outcome matching their mental model. On the other hand, text-to-image models like StableDiffusion require very specific elements and prompts to generate images that are sequentially temporal \cite{liu2022design, dehouche2023s}. To address this discrepancy caused by the lack of information given by users, our system employed GPT-4 to convert textual narratives into distinct prompts, with the number of prompts matching the number of frames specified by the designer for their storyboard. These prompts are then automatically inputted into StableDiffusion to create a complete sequence of images. We utilized task-oriented prompts for GPT-4 (see Appendix \ref{appendix_1}), which means that our prompts contain detailed and specific task descriptions. We intentionally allow for slight hallucinations in its outputs, in order to fill in the information gap between users' input and the specifics needed for generating good storyboards. Below we detail our interface design, system pipeline, prompt engineering and system implementation. %We opted to allow GPT-4 to do this because users often provided limited information when inputting their textual narrative. This caused issues as the limited information disrupted our pipeline from text to a sequence of images, and StableDiffusion required very specific elements and prompts to generate images that are sequentially temporal \cite{liu2022design, dehouche2023s}.

\begin{figure}%[h]
  %\centering
  \includegraphics[width=\linewidth]{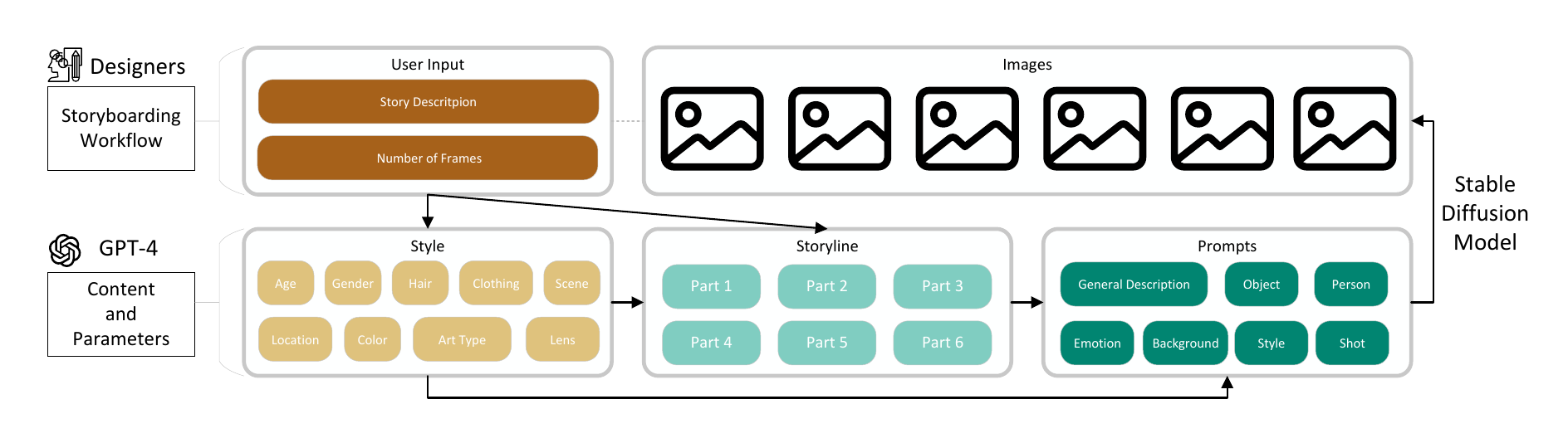}
  \caption{Parameters and Co-creation Pipeline. GPT-4 first completes the story description provided by the designer, then outputs an overarching story setting, establishing the style parameters. Subsequently, based on this setting, the story is divided into a specified number of scenes, with corresponding parameters determined. Before image generation, prompt level parameters are further added to the scenes. Once all parameters are set, the diffusion model transforms the prompts into a storyboard. Designers can oversee the entire generation process, interrupting at any stage to make modifications using natural language.}
  \Description{}
  \label{pipeline}
\end{figure}

% \begin{figure}
%   %\centering
%   \includegraphics[width=\linewidth]{Fig/HCIX-WORKFLOW-05.jpg}
%   \caption{The workflow for using StoryDiffusion.}
%   \Description{}
%   \label{WORKFLOW}
% \end{figure}

%interface
\subsection{Interface Design}

StoryDiffusion streamlines the storyboard creation process with an interface that resembles traditional storyboard templates. %It allows designers to input an idea or a narrative at any level of detail to generate comprehensive storyboards. It supports iterative refinement, meaning designers can make quick edits to the generated prompts at any stage. 
As shown in Figure \ref{WORKFLOW} and Figure \ref{pipeline}, designers begin with inputting a story description at any level of detail into the ``Story'' panel (Figure~\ref{WORKFLOW}-A). This differs from existing tools that require inputting segments of the story for each frame. Upon submitting, GPT-4 generates style elements, which can be viewed and edited under the `Style' panel (in Figure \ref{WORKFLOW}-C), that are reflective of the inputted narrative. If the generated style does not meet the designer's expectations, they can either edit the content and click ``Resubmit'' to generate new images, ``Regenerate Style'' to get new style prompt, or click ``Reset'' to clear all the style content. Refer to \textit{`Story-to-Style'} in Appendix \ref{appendix_1} for the specific prompt to accomplish this task.

After generating or modifying the overall style, clicking ``Resubmit'' will generate all the images in the number of requested frames (Figure \ref{WORKFLOW}-D). The system does not only show the images but also allow users to view and edit the generated prompts in two formats: natural language and parameterized (Figure \ref{WORKFLOW}-E). Users can switch the view of the prompt format for each image by clicking a button next to it. After any modification, clicking the regeneration button nearby will replace an image. 

\subsection{Prompting pipeline}

StoryDiffusion divides its generation process into three steps of a pipeline, as illustrated in Figure \ref{pipeline}. First is the style identification that involved GPT-4 receiving a textual narrative and automatically determining the required style from the designer's textual narrative input. This includes identifying the characters, context and the type of art needed for the storyboard. Once the style is identified, we asked GPT-4 to break the story description into six segments while keeping those style elements consistent. After creating these segments, we called GPT-4 again to create six individual prompts that are ready to be inputted into StableDiffusion \cite{automatic1111_2023, StableDiffusion2024}. 
Thus, designers do not have to think about specific prompts about visual elements. 
Each generated individual prompt contains consistent parameters to ensure the generated images align with the intended storyboard design, thus addressing the user requirements identified in our formative study. This method of prompt engineering was inspired by prior work that showed that having middle steps between each prompt led to more accurate generations \cite{chakrabarty2023spy, jeong2023zero}. Based on the findings from the formative study, we adjusted our prompt strategy to better align the images with the requirements of the UX storyboard. For instance, we highlighted the product and interface in certain images and controlled the style and tone to match the desired mood.

% To enhance the clarity, sequence, and temporal coherence of each frame in StoryDiffusion, we utilized prompt engineering. 
% One reason for the human-AI communication gap between designers' envisioned outputs and the actual GAI-generated results is because the prompts the designer's inputted did not align with what they expected the generated images to be. In StoryDiffusion, we established a framework that utilized GPT-4 to abstract away this problem. Designers can input a description of the user's journey, and if the designer wants six frames, the description is automatically transformed into six distinct prompts using a step-wise prompt engineering pipeline as shown in Figure \ref{pipeline}. Each distinct prompt is then automatically fed into StableDiffusion to generate a complete storyboard. Thus, designers do not have to think about specific prompts that they need to input into StableDiffusion for their storyboard. To create these individual prompts, we used GPT-4 to segment the textual narrative into seven essential prompt parameters in natural language (see the prompt \textit{`Story-to-Prompt} in Appendix \ref{appendix_1} to accomplish this task). These seven essential prompt parameters are listed under the `Prompts' section of Figure \ref{pipeline}. 

\subsection{Parameterization}
To create the prompts for individual images, we used GPT-4 to segment the textual narrative into seven essential prompt parameters in natural language (see the prompt \textit{`Story-to-Prompt} in Appendix \ref{appendix_1} to accomplish this task). These seven essential prompt parameters are listed under the `Prompts' section of Figure \ref{pipeline}. 
We chose the parameters based on findings from our formative study, on what elements in storyboards are important to be kept consistent and accurate to its description throughout the generation process. The first parameter is the \textit{general description}, which involves summarizing the storyline of the textual narrative into a single sentence. This helps simplify the prompt, so that only the most critical texts are inputted into StableDiffusion. The second parameter is the identification of specific \textit{objects}. This represents the different objects required based on the designer's textual narrative input, ensuring that there are specific items that exist inside the frames. The third and fourth parameters are \textit{person} and \textit{emotions}. These parameters are designed to capture the human element of the storyboard, detailing who is in the scene, what they are doing, and what they are feeling. Finally, we have the \textit{background}, \textit{style}, and \textit{shot} parameters. These prompt parameters help StableDiffusion identify what environment, artistic style, and camera angle are necessary for their generated image. 

For example, if a designer wants a storyboard with six frames, the story description is first segmented into specific styles and separate storylines using GPT-4. The identified styles and separate storylines are then combined together to create six individual prompts. These prompts are structured in a specific way to ensure that they are easily interpreted by StableDiffusion. Each prompt will also contain prompt parameters that are generated to be aligned with the designer's story description, so each individual prompt will be fed into StableDiffusion to generate frames that are visually consistent with each other. If the designer is not satisfied with the outcome of the storyboard, they can modify each prompt frame-by-frame, as shown in Step 6 of Figure \ref{WORKFLOW}, to align the generated storyboard with their expectation. 

\subsection{System implementation}

We implemented the StoryDiffusion full-stack system with a Python Flask and Vue Flow \cite{Vuejs}. The system runs on a Windows machine equipped with an Nvidia GeForce RTX 4070 GPU, which takes about one second to generate a 512x512 image with Stable Diffusion web UI's api mode and 
Deliberate-v2 model \cite{automatic1111_2023, DeliberateXpucT}. We access the OpenAI API \cite{OpenAIAPI2024} to use the GPT-4 model. 

\begin{figure*}
  %\centering
  \includegraphics[width=0.93\linewidth]{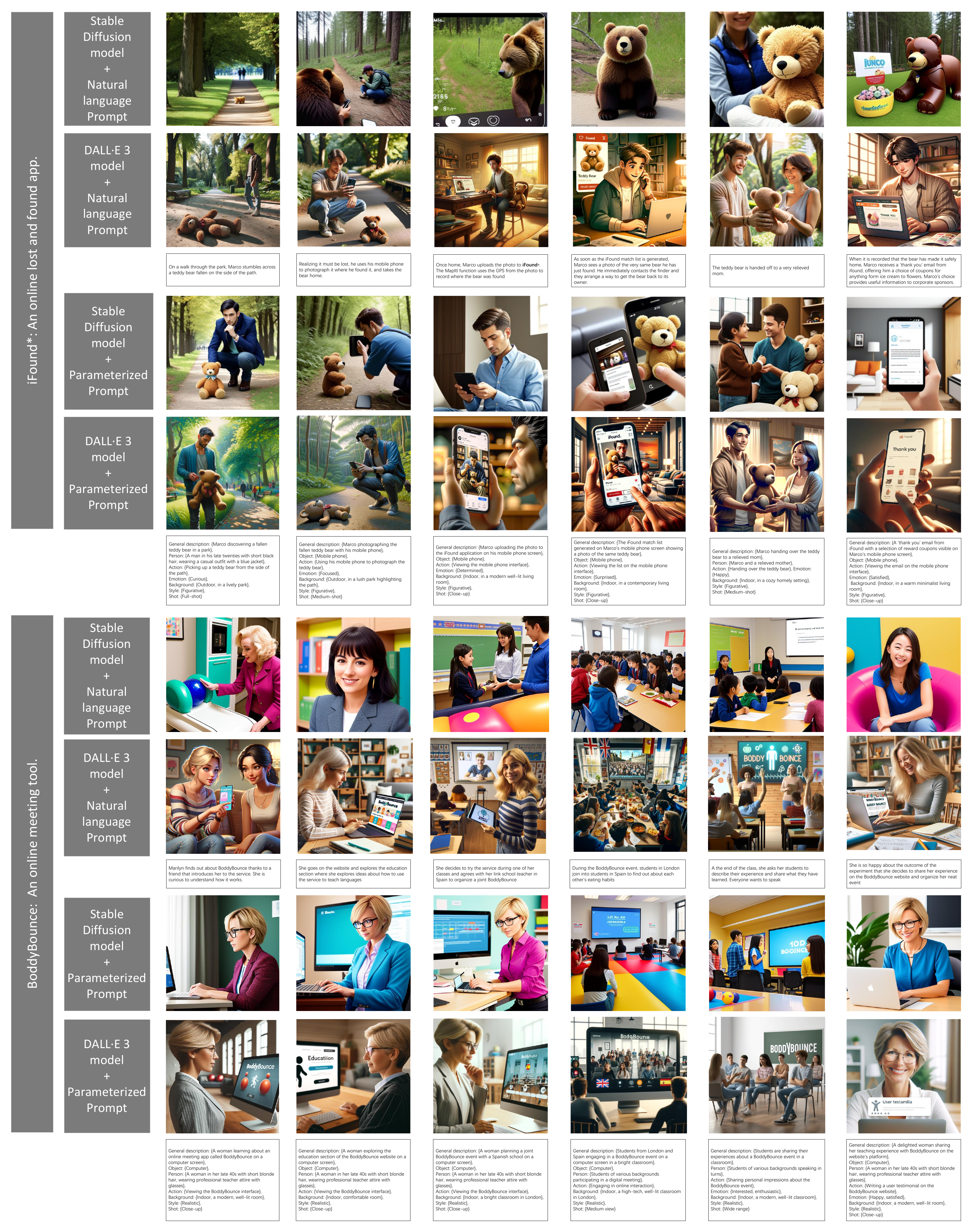}
  \caption{Test of StoryDiffusion on two stories. We conducted tests on two UX stories, with the models and prompts for each line indicated in the figure. Natural language prompts were generated using the method shown in the formative study, while parameterized prompts were generated using the StoryDiffusion system. All images generated by the Stable Diffusion model used the same model. The images processed through our system exhibited richer details, more prominent themes, and stronger coherence across both models.}
  \Description{}
  \label{EVALUATION}
\end{figure*}
\subsection{System Test}

To demonstrate the system's capability, we re-created two storyboards sourced from \cite{UXForTheMasses2024}. The results are presented in Figure \ref{EVALUATION}, with the models and prompts used for each storyboard indicated in the figure. Natural language prompts were generated using the method shown in the formative study, while parameterized prompts were generated using the StoryDiffusion system. Our comparative analysis focuses on demonstrating the effectiveness of using our prompt engineering method compared to merely inputting descriptions of the frames into Stable Diffusion to generate the storyboard, as well as the generalizability of this approach. 
We also tested the effectiveness of our method on the DALL·E 3 model. We used the Stable Diffusion model within StoryDiffusion because, despite the higher image quality of DALL·E 3, it incurs a delay of approximately 15 seconds per image, whereas the locally deployed Stable Diffusion model generates an image in less than 1 second. The high latency of DALL·E 3 would negatively impact the storyboard creation experience, which requires generating a large number of images.
To ensure a fair comparison, we restricted the regeneration process for all storyboard generation methods to a maximum of two attempts. Additionally, the textual narrative provided to StoryDiffusion was created by combining the descriptions of all the frames into a single text.

% I do not think this is necessary
% \textbf{Accuracy in converting natural language into prompts.} In natural language narrations of stories, the syntax is often complex, containing an excess of keywords and numerous thoughts of characters, posing obstacles to image generation. Through the system's transformation, the story is converted into scene descriptions with distilled main keywords, enhancing the distinctiveness of the generated images. However, during this conversion process, the system occasionally lacks detailed descriptions of scene details. For instance, the prompt for the second image in the second row lacks the 'person' parameter, leading to an inability to determine the character's appearance.

Through the system's transformation, the story is converted into scene descriptions with distilled main keywords, enhancing the distinctiveness of the generated images.
Our comparative analysis on two models, as shown in Figure \ref{EVALUATION}, shows that guiding both Stable Diffusion model and DALL·E 3 model using our prompt engineering method led to more consistency across the visuals in each storyboard frame. Specifically, we observed that when we only inputted a description of the frame (natural language prompt), the characters in the frames varied significantly across each frame. In contrast, by employing StoryDiffusion's prompt engineering technique (parameterized prompt), we achieved more uniformity and consistency in the characters, style, and angle across the storyboard without human intervention.
However, in a few cases, the absence of key parameters can lead to unstable results. The results obtained using DALL·E 3 exhibited higher image quality and fewer errors, with the ability to display keywords from the prompts within the images. However, from a storyboard perspective, the overall expressive effect of the two models was comparable when using parameterized prompts. Notably, DALL·E 3 has an integrated LLM for prompt optimization, which significantly improved the storyboard results with natural language prompts.

% also do not think this is necessary
% \textbf{Consistency across images.} 
% The samples demonstrate that compared to direct generation, our system significantly enhances the consistency and continuity among multiple images of the same story, primarily evident in the uniformity of characters, objects, and image style. This is largely due to our system's use of uniform parameters to constrain prompts, ensuring consistency in aspects like person, object, and style. However, in a few cases, the absence of key parameters can lead to unstable results, as seen with the inconsistency in clothing in the second image of the fourth row. Additionally, due to the inherent limitations of the text-to-image model, the generated characters and objects cannot be completely identical.

\section{User Study with StoryDiffusion}
\label{user_study_design}
We conducted a study with StoryDiffusion to understand the effectiveness of this tool and how it can be integrated into designers' workflows.
According to our formative study, as suggested by scenarios 1 and 2, storyboards can be used in various stages of the design process, with the most prominent being early ideation and design functionality exposition process. Therefore, we devised two types of design tasks, Concept Ideation and Concept Illustration, to explore the designer's strategies and experiences associated with our tool usage at different stages of the design process.

\begin{figure*}
  %\centering
  \includegraphics[width=\linewidth]{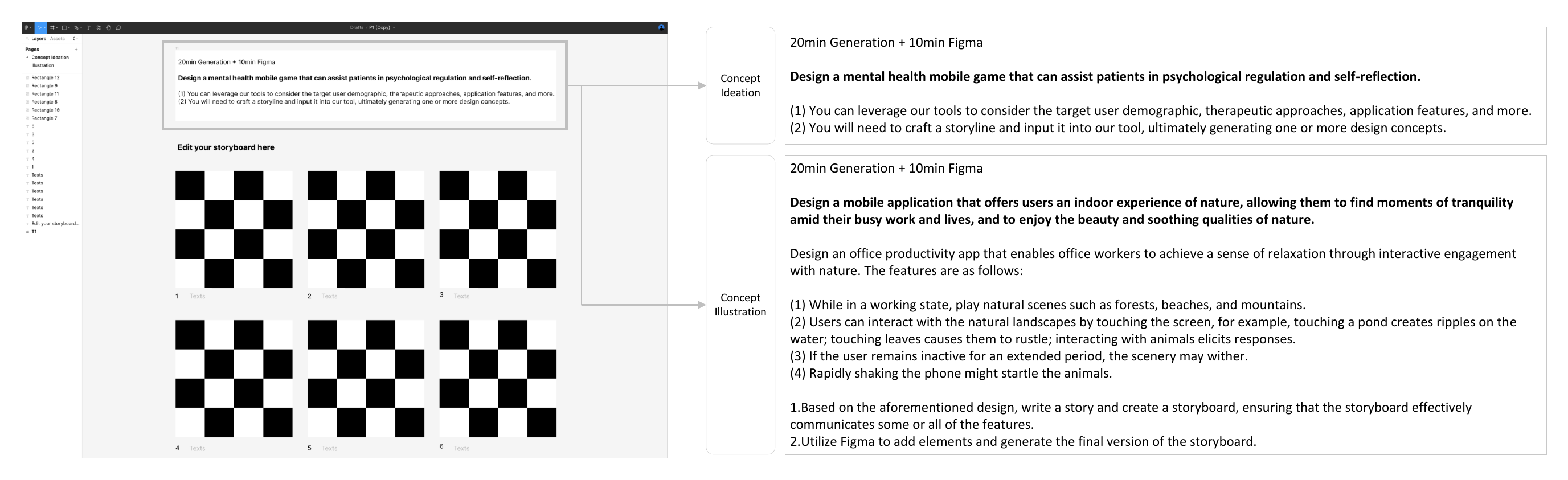}
  \caption{Figma template of storyboard. In each task, participants first read the task requirements, then used our system for creation, and finally copied the images into Figma and completed the captions.}
  \Description{}
  \label{FIGMA}
\end{figure*}

We established distinct design tasks for Concept Ideation and Concept Illustration, with the Concept Ideation tasks featuring a more general and concise design theme to allow for greater imaginative freedom, while the Concept Illustration tasks presented an articulated functional description of a product. To minimize task contingencies and learning effects, the design task themes for both Concept Ideation and Concept Illustration were dissimilar, and an additional theme was designed for each type, as depicted in Appendix \ref{appendix_2}. We named the first combination of the Concept Ideation task and Concept Illustration as Task (a), and the other one as Task (b). All four design tasks underwent meticulous discussion and consideration by the researchers to ensure that the design tasks within the same type had a similar level of difficulty. Furthermore, to explore the potential integration of StoryDiffusion with image editing tools, we chose to utilize Figma \footnote{https://www.figma.com/} for the refinement of storyboard details. Within Figma, we established templates for storyboard creation (see Figure \ref{FIGMA}), enabling participants to efficiently carry out post-processing of the images into storyboards. Before each experiment, we preset Task (a) or Task (b) for the Concept Ideation and Concept Illustration.

\subsection{Participants}
We recruited 12 participants (6 males and 6 females) aged between 18 and 26 (M=21.42, SE=0.71) from a local university through social networks. Based on variations in design experience, there were six design students at the graduate-level or higher, and we identified them as P1, P2,... P6. The other six were novice design students with less than two years of design experience, identified as P7, P8... P12. They all possessed basic storyboarding skills and have fundamental skills in design-related software. These participants voluntarily engaged in the experimental process and were permitted to withdraw at any stage of the experiment. Participants who completed the experiment received compensation of approximately \$14. The study received the research ethics approval from the Beijing Institute of Technology.

\begin{table*}[]
\caption{UX experience and Storyboarding experience of participants.}
\begin{tabular}{llll}
\hline
Participant & Education level & UX experience             & Storyboarding experience \\ \hline
P1          & Graduate        & 4.5 years and 8 projects  & 4.5 years                \\
P2          & Graduate        & 1 year and 4 projects     & 1 year                   \\
P3          & Graduate        & 3 years and 5 projects    & 3 years                  \\
P4          & Graduate        & 7.5 years and 15 projects & 7.5 years                \\
P5          & Graduate        & 2 years and 2 projects    & 2 years                  \\
P6          & Graduate        & 1 year and 1 project      & 0 year                   \\
P7          & Undergraduate   & 1.5 years and 0 project   & 1.5 years                \\
P8          & Undergraduate   & 0.5 year and 2 projects   & 0.5 year                 \\
P9          & Undergraduate   & 0.5 year and 0 project    & 0.5 year                 \\
P10         & Undergraduate   & 0.5 year and 2 projects   & 0.5 year                 \\
P11         & Undergraduate   & 0.5 year and 2 projects   & 1 year                   \\
P12         & Undergraduate   & 1 year and 1 project      & 2.5 years                \\ \hline
\end{tabular}
\end{table*}

%\subsection{Apparatus}
%Our experiments were conducted in a laboratory setting, where we established a testing environment, so the designers could access the system through desktop computers. Microphones were set up on the desktops to capture audio data during the experimental process.

\subsection{Procedure}
Before the experiment, each participant signed an informed consent form and was provided with a 10-minute introduction to familiarize themselves with generating and editing storyboards with the provided tools (StoryDiffusion and Figma). Then they started the Concept Ideation and Concept Illustration tasks. At the beginning of each task, participants were briefed on the creative objectives of the storyboard and subsequently executed the storyboard creation through StoryDiffusion. Participants had the freedom to edit the content of the storyline, modify the overall style, or adjust the story prompts corresponding to each image to align with their expectations. Following the generation of storyboard images, participants were required to import them into a Figma template for post-processing and final refinement. Each task took about half an hour. After completing both tasks, they went through a semi-structured interview for about half an hour. Each experiment lasted approximately 1.5 hours. 

\subsection{Data Collection}
All experiment sessions were recorded through video recordings and note-taking for subsequent data analysis. We collected participants' think-aloud data, where they were asked to verbalize their thoughts aloud throughout the entire experimental process. We also gathered user feedback through semi-structured interviews. The interview primarily explored inquiries across four dimensions, encompassing the following: 1) Comparing storyboards with existing tools or drawing methods, and assessing the workflow changes brought about by the introduction of StoryDiffusion. 2) Evaluating tool usage in two task processes, contrasting the different support provided by StoryDiffusion, and scoring the resulting storyboards. 3) Examining image modification methods, asking participants to provide feedback and recommendations regarding the editing features in conjunction with think-aloud data. 4) Gauging the user acceptance, with participants combining real-world usage scenarios to elucidate their willingness and reasons for using StoryDiffusion. The interviews were transcribed for qualitative analysis. %All the qualitative data above were transcribed into files with the labels of the participants.

%% \begin{table}[h]
%% \centering
%% \begin{tabular*}{\linewidth}{@{\extracolsep{\fill}}lccc}
%% \hline
%% Measures & Concept Ideation & Concept Illustration & Post-study \\
%% \hline
%% \textbf{User Experience} &   &   &  \\
%% Think-aloud Data & \checkmark & \checkmark &  \\
%% \textbf{Interaction} &   &   &  \\
%% System logs & \checkmark & \checkmark &  \\
%% Observation Recordings & \checkmark & \checkmark &  \\
%% \textbf{Storyboards} & \checkmark & \checkmark &  \\
%% \textbf{Interview} &   &   & \checkmark \\

%% \hline
%% \end{tabular*}
%% \caption{Data collected from the study.}
%% \label{Collection}
%% \end{table}

\subsection{Data analysis}
We performed inductive coding and thematic analysis \cite{thomas2006general} using NVivo, with research team members forming various themes through multiple reviews and discussions. In the next section, we present these summarized themes.  We also report quotes to support our findings, which are retained in original punctuation from the interviews but are translated from Chinese into English. 

\section{Results}
%We analyzed the logs for content modification quantitatively across the conditions. 
Based on the thematic analysis of user interviews and think-aloud data, we analyzed how StoryDiffusion assisted designers in creating storyboards by meeting their needs. Below we report the qualitative results by categories of the findings. Figure \ref{TIMELINE} was created based on the coded video recording to visualize the participants' workflows. 

\subsection{Identified Differences}

\label{strategies_employed}
\subsubsection{User-directed vs. AI-directed Creative Strategies}%The Influence of Dominant Roles on Usage Strategies.} 
% Overall, based on the dominance of either humans or AI in the creation of storyboards, participants exhibited two distinct usage strategies, namely user-directed and AI-directed storyboard generation.

In the context of storyboard creation, we differentiated two main types of creation strategies based on the observed user interactions with StoryDiffusion co-creation, namely, \emph{user-directed} and \emph{AI-directed} storyboard generation. This classification was based on whether users' initial ideas played a dominant role in the workflow of creating storyboards. Nine participants were observed to adopt a user-directed strategy. They had concrete scenarios, characters, and storylines in mind before using StoryDiffusion, so they diverged and created images that aligned with these concrete ideas. In contrast, three participants (P3, P6, P10) were observed to adopt an AI-directed strategy for both Concept Ideation and Concept Illustration tasks, who delegated the design of the specifics to the system. They tended to input a brief and abstract story to generate images directly, and then reviewed them while making minor adjustments to both the story and image illustration. In particular P10, who initially entered a story, then chose to rewrite the corresponding narrative based on the generated images. 

\begin{figure*}
  %\centering
  \includegraphics[width=\linewidth]{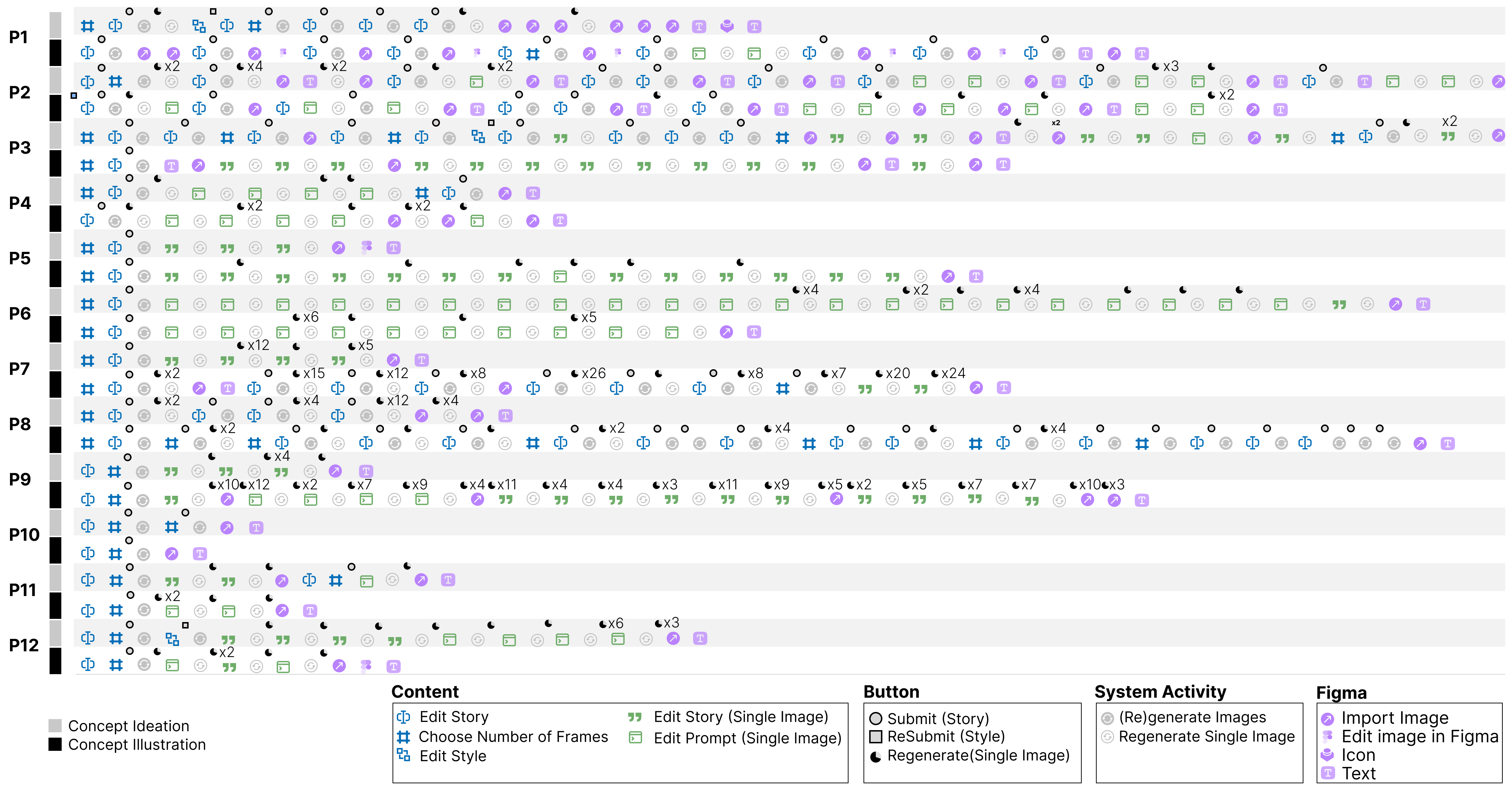}
  \caption{Usage records. We combined video recordings and system logs to summarize the timeline of content modifications and system operations by participants during the experiment.}
  \Description{}
  \label{TIMELINE}
\end{figure*}

These two strategies resulted in distinct usage patterns of StoryDiffusion. User-directed generation demanded higher precision in image representation. They would proactively provide more instructions to align the images with their vision, such as composing long and comprehensive stories or even segmenting and numbering input stories to ensure that each image represents specific content (P5, P9). Additionally, P1 and P2 framed the generated image content by inputting elements rather than complete sentences in the story input box, such as "no man, glaciers and snowfields" (P1) and "a person in the office using a cell phone" (P2). Some of these participants exhibited usage patterns distinct from the workflow of StoryDiffusion (P1, P2, P7, P8), such as inputting only a single sentence in the Story panel and generating multiple images at once to select one of them for the storyboard. Then they moved on to input the next sentence for generating the next image. As shown in Figure \ref{TIMELINE}, P8 employed this approach in both of the tasks, explaining it as follows: "Sometimes, I write a sentence and then choose to generate two images. These two images may have their respective emphases. I need one of the emphases more, so I select the one that represents it better." This highlighted the necessity of providing visual alternatives for each image. 

With an AI-directed generation strategy, the participants expressed their reliance on AI for various reasons, including its ability to generate different scenarios that facilitated the ideation (P3), the focus on story logic rather than detailed visual design (P6), and heavy usage of the captions to describe details rather than only using images (P10). This translated into a lower requirement for precision in image generation. For instance, P6, after initially generating storyboards using brief stories, relied almost exclusively on modifying individual images' prompts to create the final storyboard. P3 initially modified images by adjusting the story but later relied solely on brief modifications to the parts of the story corresponding to individual images. P10 made minimal modifications to the images and emphasized using the textual description in the captions to convey information, with images serving as supplementary elements.

% He explained, "Firstly, I already had a rough idea of what each scene would look like. So, I felt that going image by image would allow for better alignment. Actually, I can also try multiple images, but with multiple images, I can only control the overall story, and the system generates these scenes on its own, and I'm worried it won't align well." 

%\subsubsection{Novice Designers vs. Expert Designers}
%\label{Novice_Expert Designers}

\subsubsection{Concept Ideation vs. Concept Illustration}%The Influence of Design Phase on Usage Strategies} 
Due to the distinct purposes of storyboard creation in Concept Ideation and Concept Illustration, different usage strategies emerged among participants. Overall, participants perceived that StoryDiffusion primarily served as an inspiration catalyst for Concept Ideation (P1, P3, P5, P6, P8) and as a means to visualize ideas for Concept Illustration (P1, P2, P5, P8, P9). The differing roles attributed to StoryDiffusion led participants to place relatively more emphasis on story logic and image precision in Concept Illustration. For instance, P9 stated, "In Concept Ideation, I require that the images should reach a certain level of accuracy, and I don't have very high overall demands. However, in Concept Illustration, my requirements become more complex. For example, I need the viewer to immediately perceive the characters' psychological states or the ambiance." 

We also asked each designer to rate the quality of the generated storyboards on a scale of 1 to 10 and provide reasons for their ratings. The results showed that participants gave an average score of 6.67 for Concept Ideation and 6.88 for Concept Illustration. For Concept Ideation, the main reasons for the storyboard's score were the provision of more creative ideas (P1, P7, P10), while deduction points were attributed to continuity issues (P3, P6, P11, P12) and lower image quality (P2, P9, P12). In the case of Concept Illustration, the primary factors contributing to the storyboard's score were participants expressing that more detailed descriptions led to higher image quality (P3, P4, P11, P12) and precision (P1, P2, P12). Deduction points were related to the difficulty of StoryDiffusion in understanding and presenting specific descriptions, as indicated by P3, "It seems like it doesn't understand how to represent `shaking the phone,' so I'm not very satisfied with this image." The results indicated that although participants demanded greater control over details in the Concept Illustration task in usage strategy, the resulting more precise storyboards also contributed to higher participant satisfaction.

\subsection{Image Editing Strategies}
\label{influence_designer}
In this section, we detailed the specific editing methods that participants used to modify images with the current solutions provided by StoryDiffusion. 

\subsubsection{Enhancing Precision}
Regarding precision requirements, participants faced visual demands primarily related to the composition, style, elements, characters, and scenes. For composition, participants focused on modifying the visual focal point (P5) and having control over the arrangement of elements (P7), meaning greater control over the positioning and angles of elements and camera shots. For example, P7 changed the prompt to "On the left is a little girl, and on the right is a phone screen." adjusting the composition by specifying the positions of elements. For style, P1 attempted to alter the color parameter of the Style to "Gloomy colors with the environment" to shift the image tone of the storyboard from bright to more somber, aligning it with the "unpleasant mood" story context. In terms of adjusting elements, participants mostly addressed the issues by modifying the prompts and stories corresponding to each image. For example, describe the lake water as "clear and transparent" (P9) or change the prompt from "Computer" to "Phone" (P5). Concerning characters, participants' modifications mainly revolved around the number of characters, actions, facial expressions, and gaze. Most of these adjustments could be made by editing the prompts and stories of the target images. However, certain editing operations required relatively more effort because corresponding functionalities were not provided. For example, P1 wanted scenes without human beings, but StoryDiffusion defaulted to including characters. As a result, P1 spent time removing descriptions related to people. Regarding scene adjustments, they easily resolved the background transitions by adjusting the prompts or regenerating the image (P5, P9). 

Participants introduced new requirements based on their precision criteria. These approaches involve enhancing control over style and image details, such as providing filter functionality (P1, P2). P5 further expressed a desire to enhance the model's understanding of specialized design terminology, such as "avatar". P6 expressed a need for control over element states, such as controlling the degree of water surface ripples. P1, based on his experience adjusting the display of characters in the images, expressed a desire for an option in the initial phase to specify whether characters should appear. Moreover, P3 and P9 proposed the addition of an image reference feature, allowing users to upload reference images before creating storyboards to provide StoryDiffusion with a generation reference.

\subsubsection{Improving Continuity}
Concerning continuity requirements, participants edited their work to address the following issues: 1) Continuity of characters, which includes uniformity in the appearance, clothing, and number of characters. 2) Continuity of scenes, including consistency in background items and the overall setting. The continuity of characters and scenes was improved by modifying prompts and style. For example, P12 maintained the consistent appearance of characters in a "blue shirt" by modifying the style. 3) The logical flow of the story, encompassing its coherence and logic. Participants made modifications directly to the images. For example, P1 and P12, when confronted with two images with minimal differences in story content, opted to delete one to streamline the narrative. P5 and P10 adjusted the sequence of images to enhance the overall coherence of the story.

Participants mentioned other functional requirements, encompassing improvements in interface design and manipulation of image elements. For instance, P4 and P5 wished for greater flexibility in image editing to facilitate partial redrawing. P5 suggested addressing continuity in backgrounds by enabling image layering, allowing for changes to characters and design elements while keeping the background unchanged.

\subsubsection{Story Modification vs. Prompts Modification}
% Due to the varying degrees of control, we categorized the textual modifications into two groups: overall modification and partial modification. Overall modification refers to editing the story or adjusting the style content to regenerate all the images at once. Partial modification involves adjusting a specific story or prompt corresponding to a single image. Through this categorization, we aimed to examine the differences in participants' control over image editing.
Figure \ref{TIMELINE} indicates that most participants employed a combination of overall modification and partial modification methods to enhance their storyboards. This phenomenon helps illustrate that the integrated storyboard creation process, combining text-to-text and text-to-image generation approaches, was reasonably well-grasped by the participants. We also observed some participants adapting their textual narrative or captions in the storyboard based on or inspired by the generated images. Only a couple of participants solely modified the textual narrative or the image prompts. P6 solely edited the prompts, emphasizing the consistency of the story. He adjusted the prompts to ensure uniformity in scenes and object content among the images. P5 exclusively edited the story corresponding to each image. She mentioned that GPT-4's segmentation and elaboration of the complete story aided her in adding richer details to each frame, making her prefer modifications to the natural language descriptions.

\subsubsection{Post-processing of Images}
For the majority of participants, Figma served as a tool for storing images and adding textual descriptions. Some participants carried out additional editing of the images, such as P1, who performed cropping and added icons, and P5 and P12, who used Figma to adjust the sequence of images. Based on their post-processing of images, participants further articulated their desired functionalities, which will be discussed in the next section.

\begin{figure}
  %\centering
  \includegraphics[width=\linewidth]{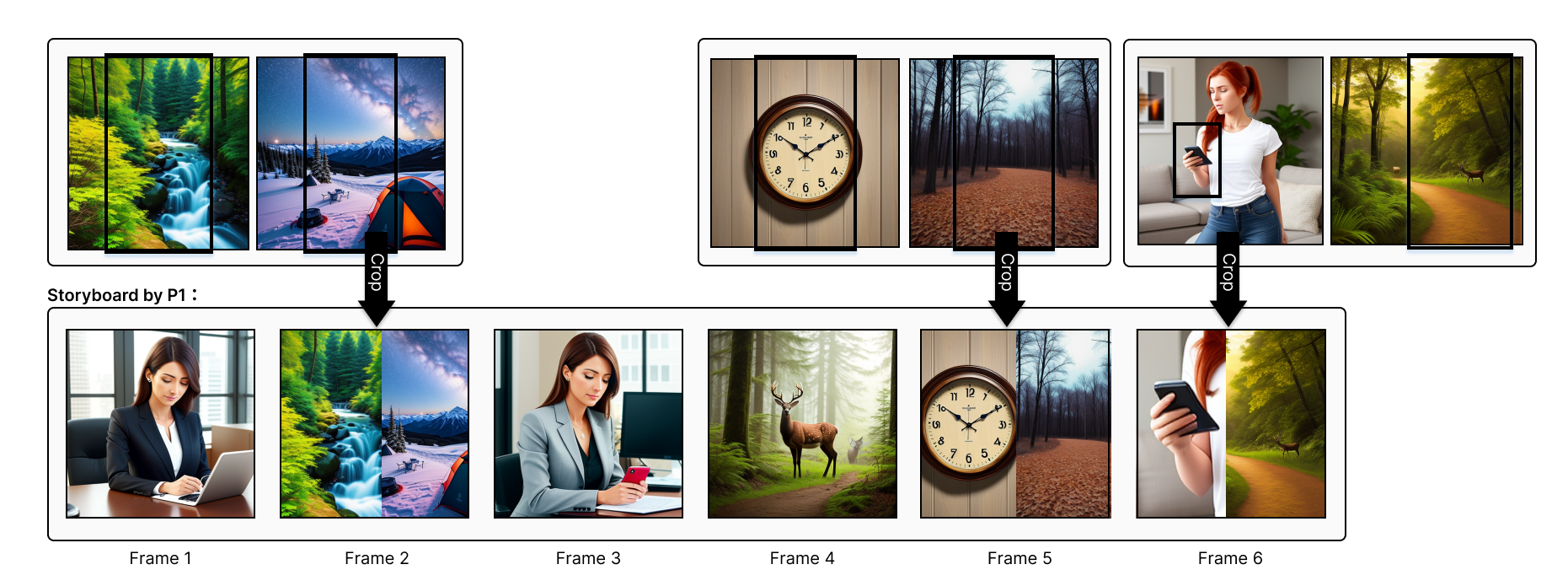}
  \caption{The Storyboard by P1. P1 aimed to display more information in the storyboard, hence performing cropping and stitching of images in Figma.}
  \Description{}
  \label{P1}
\end{figure}

Based on the functionality within Figma, participants expressed expectations for editing features in StoryDiffusion. These primarily include the addition of filters (P1, P4, P9) and adding character dialogue boxes (P12). P1 and P3 additionally proposed image splicing functionality, where, for instance, he used more than one image in frames 2, 5, and 6 to convey more content within a single frame (See Figure \ref{P1}) when creating storyboards.

\subsection{User Feedback and Acceptance}
\label{influence_stage}

\subsubsection{Compared with Existing Storyboarding Tools}
When comparing StoryDiffusion to participants' usual storyboard creation tools, primarily hand-drawing applications, participants expressed that its overall advantages include enhancing design content (P9, P11), improving storytelling (P5, P9), reducing workload (P2, P3, P6, P10, P11, P12), sparking inspiration (P3, P7), and saving time (P1, P2, P4, P6, P8, P10, P11). Participants also indicated that the unique features provided by StoryDiffusion, which assists in organizing logic and content allocation, had altered their approach to creating storyboards. For instance, P12 stated, "The usual method to create a storyboard is to plan what content should be presented on each image. But with this tool, I can just jot down my ideas, and it automatically assigns them to each image." P1 similarly appreciated the system's automatic segmentation of story content. P5 expressed, "Before, when I used to create storyboards, I would first refer to other examples. With this tool, it provides me with a direction, so I don't need to rely on other examples as much." Furthermore, the ability to make modifications based on individual images was considered quick and effective, as P6 mentioned, "The biggest advantage is that I can intuitively and instantly make adjustments to individual images." However, participants also expressed concerns about using StoryDiffusion, primarily related to the reliance on AI-enabled tools leading to complacency. For example, P6 remarked, "While it can quickly provide a lot of materials, I believe it is not a very good thing if one rely on it for every aspect of creativity." Besides, P3 pointed out that the convergence in the design presentation of images could limit the generated images' stylistic diversity, making it challenging to achieve greater creative breakthroughs.

\subsubsection{Feasibility of Practical Use}
Regarding the practical value of StoryDiffusion, participants expressed their willingness to use it in real-world applications. Their reasons included time-saving (P11, P12), aiding ideation (P4, P5, P6, P7), conveying inspiration (P2, P9), providing references (P6), and integrating it into the design process (P3). Notably, 8 out of 12 designers showed a preference for using StoryDiffusion in Concept Ideation tasks, driven by two primary reasons. Firstly, StoryDiffusion can offer more inspirational possibilities at this stage. For instance, P3 stated that this enabled him to "obtain random design outputs, which are more intriguing." P9 believed it could "stimulate creative collisions." Also, it provides greater creative freedom. Participants found that in Concept Illustration tasks they had more content to express, and it became challenging to articulate all their ideas clearly, such as participants spontaneously desiring more detailed descriptions and additional revisions to enhance image precision. For example, P8 remarked that in Concept Illustration tasks, there is "a lot of content to describe." The primary reason designers favored the Concept Illustration task cited was their inclination towards using StoryDiffusion solely as a drawing tool. For example, P12 stated, "After I have a clear idea in mind, providing specific instructions to generate images, using it solely to present my ideas would be better. Using it when I haven't fully figured out my ideas can easily get lost in an excessive amount of image information."

\section{Discussion}
\label{disc}
In this paper, we introduced StoryDiffusion, a GAI tool designed to transform textual narratives into a visual storyboard. Through our user study, we identified two main creative strategies that participants undertook: user-directed creation, where designers specify their own model of what the final outcome will look like before using StoryDiffusion, and AI-directed creation, where StoryDiffusion steered the designers to the final storyboard outcome. 

We also compared the use of StoryDiffusion for concept ideation and concept illustration tasks, highlighting distinct preferences for using StoryDiffusion in both cases. In particular, we noted some particular prompt engineering strategies that designers employed to generate precise images while maintaining continuity across the storyboard. In addition, we noted 
% that designers generally preferred using StoryDiffusion for concept ideation tasks over concept illustration tasks, indicating 
that StoryDiffusion can facilitate ideation via the generation of inspirational stimuli and alternative design ideas. %Participants overall appreciated StoryDiffusion for giving them the ability to focus more on creation and ideation rather than illustration when using our system over traditional methods of drafting storyboards (i.e. pen and paper). 
Despite these advantages, participants also raised concerns about the potential for designers to become overly reliant on AI for creativity, which could hinder creative breakthrough. In the rest of this section, we discussed the implications of our findings on future designs of GAI storyboard systems.

\subsection{AI-Directed and User-Directed Storyboard Generation}
\label{disc:ai-user}
From our user study, we identified two distinct groups of participants that engaged with StoryDiffusion: user-directed and AI-directed. User-directed designers exhibited a preference for high precision in the generated images, seeking to closely align the AI generation with their own preferences. Conversely, AI-directed designers were more flexible with whatever StoryDiffusion generated, allowing them to concentrate on other facets of the design task, such as ideation and creation. Notably, the majority of our participants fell into the user-directed category.

This observation potentially challenges some popular assumptions about the ideal use of GAI in design. Previously, it had been suggested that GAI would be most useful for designers by helping them explore new ideas, finding inspirations, and customizing designs quickly \cite{lu2022bridging}. This would imply that a certain level of randomness in the generated outputs is desirable and beneficial. However, our findings, as explained in Section \ref{strategies_employed} and as illustrated in Figure \ref{TIMELINE}, revealed that participants often repetitively iterated prompts to achieve a specific storyboard output. This process can be inconsistent and tedious, which underscores the limitations with using GAI for design tasks in that their outputs are not easily controllable and frequently random. As a result, we argue that there is a need for a generative storyboard system that strikes an optimal balance between generation precision (to align with the designer's narrative vision) and maintaining some level of randomness (to foster ideation and creativity). While previous research has addressed similar challenges by training new variants of AI models \cite{mozaffari2022ganspiration}, such an approach that requires retraining new AI models to address this issue is not feasible for large pre-trained text-to-image models like StableDiffusion. Thus, we believe that one potential solution to this is exploring how we can support designers through enhanced prompt strategies and recommendations to designers in controlling the randomness of large pre-trained text-to-image models. Such system for enhanced prompt recommendations has shown promise in prior work \cite{almeda2023prompting}, so we envision that the integration of similar methodologies into systems like StoryDiffusion can offer a promising direction for supporting designers. 

% Furthermore, %our results imply that advancements in this area could be particularly beneficial for novice designers. R
% results from our user study suggest that experienced designers were more creative and resourceful in making use of generated images that might not be perfectly aligned. %formulating effective prompts to realize their desired storyboards. Most notably, 
% Whereas novice designers often resorted to more basic forms of interactions, such as repeatedly clicking the ``regenerate'' button, in their effort to achieve the intended results. 
% While generating perfect images might always be a challenge, this observation may inspire future GAI systems to support user appropriation in other ways than merely focused on alignment. 
%Based on these insights, we believe that this vision could pave the way for the development of a GAI storyboard system that significantly improves both expert and novice designers in integrating this tool into their design process.

\subsection{Concept Ideation vs. Concept Illustration}
\label{disc:concept}
Our user study revealed a preference amongst participants for employing StoryDiffusion in concept ideation tasks rather than for concept illustration purposes. This preference is attributed to StoryDiffusion's capability to rapidly generate a wide array of design content, therefore offering a richer pool of inspirational stimuli for concept ideation. In contrast, concept illustration tasks demanded highly specific visuals, requiring designers to make numerous iterations and revisions when using StoryDiffusion to generate the storyboards. 

Previous research on utilizing GAI for design tasks has demonstrated the utility of GAI in supporting designers across both result-oriented and process-oriented tasks, with concept ideation aligning more closely with process-oriented tasks and concept illustration with result-oriented tasks \cite{verheijden2023collaborative}. Our findings extend on this body of work by illustrating a general preference amongst designers for leveraging GAI in tasks that are more process-oriented rather than result-oriented. We speculate that designers generally preferred using our tool for concept ideation over concept illustration because of StoryDiffusion's capabilities to create an entire storyboard based on one narrative, which can be repeatedly regenerated, giving designers more freedom to explore different ideas. Recognizing these task-specific user requirements for GAI helps us develop future systems to support them differently. 
%However, recall from Section \ref{disc:ai-user} that participants also had a strong desire for precision in their generation, suggesting a need for these systems to strike a balance between generating a wide range of design alternatives quickly while also allowing for the generation of specific, detailed images. 

% In light of these findings, we argue that future generative storyboard systems must navigate the trade-offs between fostering creative exploration and maintaining precision in its generated images. Further research is necessary to explore these trade-offs and to determine an optimal balance that supports the designer's needs across different design tasks. 

\subsection{How did StoryDiffusion Change the Way Designers Created Storyboards?}
\label{disc:how?}

%Storyboards are usually created via traditional methods like hand-drawing to succinctly convey the user's experience, which enables designers to rapidly visualize ideas and comprehend a user's journey \cite{truong2006storyboarding}. To do this, it is common practice for designers to first outline each plot step and the basic story structure, which involves having a clear idea of what each storyboard frame should visually depict. Following this, designers would typically flesh out each frame with detailed illustrations to capture the emotions, backgrounds, and actions of each scene. In contrast, 
Compared with the traditional manual process of drawing storyboards, StoryDiffusion has led to different creative processes. %approaches of creating storyboards. 
Designers start by inputting a textual narrative of the user experience. StoryDiffusion then generates a complete visual storyboard based on this narrative. Designers then have the flexibility to either regenerate the entire storyboard or iterate on individual frames to produce new images. Moreover, they can also opt to revise the entire narrative for the whole storyboard if they disagree with the outputted content. This highlights a unique feature of our system: integrating narrative development and image generation into a single pipeline.

% This has led to interesting user experiences for the participants in our study. 
In our study, we asked the participants how creating storyboards with StoryDiffusion has differed from traditional means of creating storyboards. We noticed that our participants generally appreciated StoryDiffusion for helping them expedite the storyboard creation process because it allowed the designers to generate a storyboard almost instantly from a complete or incomplete narrative.  %which provided them with a clear direction immediately. With this extra time and reduction in cognitive effort, designers were able to put more effort into other aspects of the storyboard like searching for new ideas, enhancing design content, and improving the storytelling. This indicates there are significant benefits from using GAI to support the creation of storyboards. 
Our findings show that most of our participants went back and forth between narrative development and image iteration, and that the generated images helped them revise their story. This confirms that the processes of narrative and visual development for storyboards are intertwined, thus should be integrated together into one tool. One potential implication for future design is to even support the reverse AI generation from user-selected images to textual narrative, thus realizing a bi-directional cross-modality story generation process. %embedded in our current pipeline. 

% Wethe story recommends that future iterations of generative storyboard systems should incorporate a feature that allows designers to create revised narratives based on their preferred images or storyboards. 

% Despite these benefits, our study also unveils critical concerns associated with an over-reliance on AI in the creative process. One notable apprehension is the potential stifling of creative capabilities and the homogenization of stylistic diversity in a storyboard if designers become too reliant on GAI to create storyboards. This apprehension echoes findings from prior research, which have indicated that AI's influence can extend to impairing decision-making and fostering a dependency that diminishes human initiative and creativity \cite{ahmad2023impact}. In light of these considerations, we advocate for future investigations into the ethical and careful integration of GAI within the design process. While prior literature has affirmed the potential of GAI to enhance the design process \cite{vinchon2023artificial}, it is imperative to explore methodologies that ensure creativity is not inadvertently constrained by an over-dependence on AI. An approach that could be considered would involve initiating design tasks without AI assistance, resorting to GAI only when designers encounter creative impasses like design fixation. This strategy aims to leverage AI as a tool primarily for inspiration and alternative design perspectives, thus fostering a balanced integration of AI in the design process.  

\subsection{Future Improvements on Generative Storyboard Systems}
\label{disc:future}

% When designing StoryDiffusion, there was a significant amount of engineering effort to maintain continuity and flow across the frames in our storyboards. However, the inherent randomness and inconsistency of text-to-text and text-to-image models posed significant challenges in replicating specific details, even with meticulous prompt engineering and manipulation. This aspect of our system led to participants bringing up several issues with our system during our user study. In particular, 
When asked about the room for improvement of StoryDiffusion, participants underscored the necessity for higher accuracy and continuity across each frame. We believe this issue will be mitigated as LLMs and text-to-image models continue to improve in the future. In fact, the recently upgraded DALL.E 3 model can already generate a visibly better outcome using the same prompts, compared to the StableDiffusion model used in our experiment, as demonstrated in Figure~\ref{EVALUATION}.
% Based on these findings, we propose that future generative storyboard systems incorporate features and interfaces designed to guide users in crafting prompts that yield visuals closely aligned with their envisioned designs. 
Moreover, as detailed in Section 6.2, participants still needed to make various modifications on the generated images. Therefore we advocate for the inclusion of advanced image modification features
% , and, specifically, we recommend the integration of tools 
that enable more refined image editing, resizing, cropping, filtering and character manipulation capabilities. %Such enhancements, we argue, will pave the way for new generative storyboard systems that can significantly improve the overall user experience. 
In addition, there were demands for expedited generation of the storyboards, so the speed of generation may not align with the participant's expectation.

\subsection{Limitations}

% Users had complaints about the controllability of StoryDiffusion, as highlighted by participant feedback, and this reported challenge led to many participants advocating for an alternative to prompt iteration for visual modification. We acknowledge this as a limitation of our system design, and in the future, we would like to create a system that incorporates more elements of control into the image modification, such as giving the designers the ability to resize images, crop images, and add filters or effects. Moreover, there were demands for expedited generation of the storyboards, so the speed of generation may not align with the participant's expectation. This discrepancy in generation speed can potentially introduce an unintended confounding variable, which should be acknowledged as a limitation within our user study's context. Furthermore, 
It is pertinent to note that our user study predominantly involved design students as participants. While these individuals may have a foundational understanding of design principles and technical competencies, their expertise does not equate to that of professional designers. In the future, we hope to study such systems with professional designers as participants to mitigate this limitation. 

%Future Work
\section{Conclusion}

We developed StoryDiffusion, a system that integrated text-to-text and text-to-image generation models to assist designers create storyboards. Different from prior works focusing on the alignment between scripts and generated images, it provides AI support for both narrative development and image creation in one tool thus allowing users to smoothly transition between them. Our user study revealed user-directed and AI-directed strategies and provided insights into users' resourcefulness in incorporating StoryDiffusion in their diverse workflows. While precision in text and image alignment is indeed important for GAI applications, we argue that supporting the entire creative process unleashes users' ability to adapt their work and make use of imperfect image generation. 
In addition, the findings indicated a slight preference for designers using our system in concept ideation tasks over illustration tasks because our system afforded them the capability to explore diverse ideas and design alternatives. 
% We also noted that incorporating StoryDiffusion into the storyboard creation process, as opposed to more traditional methods, changed how designers approached each design task. This observation suggested that using GAI in creative storyboard desig tasks could potentially come with both advantages and disadvantages. 
% Finally, we discussed some additional challenges related to using natural language for precise image generation, indicating areas for future refinement. 
These insights help pave the way for future development of GAI storyboarding tools that more effectively support designers' creative processes. 

% \begin{acks}
% This work was partially funded by the Google Faculty Research Award.
% \end{acks}

\bibliographystyle{ACMReferenceFormat}
\bibliography{main}
%TC:ignore

%\begin{comment}
\appendix
\section{System Prompt}
\label{appendix_1}
%\subsection{Story-to-Style Prompt}
\textbf{Story-to-Style Prompt:} 

You are now assuming the role of a prompt generator for a generative AI named ``Stable Diffusion.'' This AI specializes in creating images from provided prompts. Your task involves creating UI storyboard prompts based on the stories I provide. The stories I will give you will all pertain to UI design. It is crucial that you adhere to the following guidelines and refrain from altering the structure in any manner. Your objective is to generate the appropriate Storyboard style using the information I provide for the Stable Diffusion AI. If the content I provide is brief or just a hint or requirement for a story, please imagine and complete the entire story based on what I provide, but never output any stories, and then just generate styles as requested.

Prompt structure, including the following 8 parameters:

\begin{enumerate*}[label=, itemjoin={\quad}]
    \item Age:\{\}
    \item Gender:\{\}
    \item Hair:\{\}
    \item Clothing:\{\}
    \item Scene:\{\}
    \item Location:\{\}
    \item Color:\{\}
    \item Art type:\{\}
    \item Lens and Shot:\{\}
\end{enumerate*}

For example: Age:\{5-7\}, Gender:\{female\}, Hair:\{brown curl\}, Clothing:\{blue dress\}, Scene:\{under the soft glow of her desk lamp\}, Location:\{Indoor, in Cindy's warm and comfortable bedroom\}, Color:\{warm tones\}, Art type:\{realistic\}, Lens and Shot:\{Medium Shot\}

Let me provide descriptions for each parameter:

\begin{itemize}
    \item Age: Indicate the age range or specific age of the character in the scene. This provides context on the maturity and appearance of the individual.
    \item Gender: Detail the gender of the character, allowing for a clearer picture of the individual.
    \item Hair: Describe the type and color of the character's hair.
    \item Clothing: Indicate the attire of the character, contributing to their overall look and the scene's setting.
    \item Scene: Set the atmospheric mood or immediate surroundings in which the character is placed.
    \item Location: Detail the broader setting or venue where the action takes place, giving a sense of place and ambiance.
    \item Color: Specify the dominant or notable colors in the scene, helping to set the mood and visual theme.
    \item Art type: Choose the artistic style of the depiction, guiding the visualization, e.g., realistic, sketch, Disney cartoon.
    \item Lens and Shot: Pick the type of camera view, which determines how the scene is visually framed and presented.
\end{itemize}

Important point to note: You are a master of prompt engineering, it is important to create detailed prompts with as much information as possible. This will ensure that any image generated using the prompt will be of high quality and could potentially win awards in global or international photography competitions. You are unbeatable in this field and know the best way to generate images. I will provide you with keywords and you will generate only one prompt in a code cell without any explanation just the prompt like the example I provided before. This will allow me to easily copy and paste the code. Please make sure to use a realistic style in the images.

%\subsection{Story-to-Prompt Prompt}
\vspace{10pt}
\noindent\textbf{Story-to-Prompt Prompt:}

You are now assuming the role of a prompt generator for a generative AI named ``Stable Diffusion.'' This AI specializes in creating images from provided prompts. Your task involves creating UI storyboard prompts based on the stories I provide. The stories I will give you will all pertain to UI design. It is crucial that you adhere to the following guidelines and refrain from altering the structure in any manner. Your objective is to generate the appropriate Storyboard prompts using the information I provide for the Stable Diffusion AI.

Prompt structure, including the following 8 parameters:

\begin{enumerate*}[label=, itemjoin={\quad}]
    \item General description: \{\}
    \item Object: \{\}
    \item Person: \{\}
    \item Action: \{\}
    \item Emotion: \{\}
    \item Background: \{\}
    \item Style: \{\}
    \item Shot: \{\}
\end{enumerate*}

For example: General description: \{A boy playing with a dog in a park\}, Person: \{A boy with a red hat and freckles\}, Action: \{playing fetch with a golden retriever dog\}, Background: \{outdoor, a sunny park with a lake\}, Shot: \{close-up\}

Make sure you generate PIC-NUM-NEEDED prompts in this format. It is important to know that you do not need to add all 8 parameters in a prompt every time; you can generate 7 or fewer parameters in a prompt according to your needs. Word order and effective adjectives matter in the prompt.

Let me provide descriptions for each parameter:
\begin{itemize}
    \item General description: Describe the core information of the image scene. Reflect the core of what needs to be shown, especially when the image size is limited.
    \item Object: Detail the main object.
    \item People: Describe the appearance and attire of the characters.
    \item Action: Describe the actions of the people.
    \item Emotion: Describe the facial expressions of the characters.
    \item Background: Specify the scene setting.
    \item Style: Choose between abstract or figurative style.
    \item Shot: Choose the type of camera view.
\end{itemize}

Curly brackets are necessary for the prompt to provide specific details about the subject and action. These details are important for generating a high-quality image.

Important to note: The prompts you furnish will be in English. I will provide a comprehensive story and 'styles' of pictures divided by ``//''. You will divide this into corresponding PIC-NUM-NEEDED parts and generate prompts for each part to create a coherent sequence of prompts. Consistency must be maintained among prompts based on the 'style' information.

You are expected to generate PIC-NUM-NEEDED prompts within a code cell, without any additional explanations, solely providing the prompts. This streamlined approach will facilitate easy copying and pasting of the code.

\section{Tasks in Concept Ideation and Concept Illustration}
\label{appendix_2}
\noindent\textbf{Task (a)}\\
\noindent\textbf{Concept Ideation:} 

\noindent\textit{Design a mental health mobile game that can assist patients in psychological regulation and self-reflection.}

(1) You can leverage our tools to consider the target user demographic, therapeutic approaches, application features, and more.
(2) You will need to craft a storyline and input it into our tool, ultimately generating one or more design concepts.

\vspace{10pt}
\noindent\textbf{Concept Illustration:}

\noindent\textit{Design a mobile application that offers users an indoor experience of nature, allowing them to find moments of tranquility amid their busy work and lives, and to enjoy the beauty and soothing qualities of nature.}

Design an office productivity app that enables office workers to achieve a sense of relaxation through interactive engagement with nature. The features are as follows:
(1) While in a working state, play natural scenes such as forests, beaches, and mountains.
(2) Users can interact with the natural landscapes by touching the screen, for example, touching a pond creates ripples on the water; touching leaves causes them to rustle; interacting with animals elicits responses.
(3) If the user remains inactive for an extended period, the scenery may wither.
(4) Rapidly shaking the phone might startle the animals.
1. Based on the aforementioned design, write a story and create a storyboard, ensuring that the storyboard effectively communicates some or all of the features.
2. Utilize Figma to add elements and generate the final version of the storyboard.

\vspace{10pt}
\noindent\textbf{Task (b)} 

\noindent\textbf{Concept Ideation:}

\noindent\textit{Design a mobile application that offers users an indoor experience of nature, allowing them to find moments of tranquility amid their busy work and lives, and to enjoy the beauty and soothing qualities of nature.}

\noindent(1) You can leverage our tools to consider the target user demographic, therapeutic approaches, application features, and more.
\noindent(2) You will need to craft a storyline and input it into our tool, ultimately generating one or more design concepts.

\vspace{10pt}
\noindent\textbf{Concept Illustration:}

\noindent\textit{Design a mental health mobile game that can assist patients in psychological regulation and self-reflection.}

Designed an AR-based mental health mobile game that helps individuals confront their inner selves through interactive experiences. The features are as follows:
(1) Users can select a comfortable small space, like a bedroom, to launch the AR game. The camera scans the environment to generate different storylines and tasks. Different spaces create distinct tasks and storylines.
(2) Users need to complete various AR tasks, such as engaging in a conversation with a virtual character reading a book at a virtual desk, immersing themselves in a third-person perspective to experience their life and emotions.
(3) Different choices within tasks yield different emotional elements, such as happiness, calmness, and courage. Collecting these elements allows users to piece together their psychological state.
(4) A mood diary is provided, allowing users to record their emotional states. This feature enables reflection and analysis of their own emotions.
1. Based on the aforementioned design, write a story and create a storyboard, ensuring that the storyboard effectively communicates some or all of the features.
2. Utilize Figma to add elements and generate the final version of the storyboard.

\section{Participants' experience outlines}
\label{appendix_3}
\textbf{P1:} Concept ideation: P1 edits \emph{story} to depict each image in the storyboard, rather than generating the entire storyboard at once. P1 considers that precise control over each image can accurately depict the scenarios, processes, and outcomes of using the product. Concept Illustration: P1 edits \emph{story} to depict each image in the storyboard, rather than generating the entire storyboard at once. P1 considers that precise control over each image should align the images with key elements of the design illustration.

\noindent\textbf{P2:} Concept ideation: P2 edits \emph{story} to depict each image in the storyboard, rather than generating the entire storyboard at once. P2 considers that precise control over each image can accurately depict the scenarios, processes, and outcomes of using the product. Concept Illustration: P2 edits \emph{story} to depict each image in the storyboard, rather than generating the entire storyboard at once. P2 takes precise control over each image to illustrate the UI and usage of the product.

\noindent\textbf{P3:} Concept ideation: P3 edits simple sentences in \emph{story} to conclude the storyboard she wants and rethink based on the AI-generated results from these comprehensive descriptions. P3 co-works with AI to iterate and present the logical and rigorous process of product usage. Concept Illustration: After editing \emph{story} to conclude the storyboard she wants, P3 edits \emph{story of single image} to take precise revision over each image. P3 converges her idea to illustrate the UI and usage of the product.

\noindent\textbf{P4:} Concept ideation: P4 edits \emph{story} to generate the entire storyboard at once and adjusts \emph{story of single image} to revise images that do not match her expectations. P4 presents the storyboard to describe the scenarios, processes, and outcomes of using the product. Concept Illustration: Same as above.

\noindent\textbf{P5:} Concept ideation: P5 edits \emph{story} to generate the entire storyboard at once and adjusts \emph{story of single image} to revise images that do not match her expectations. P5 presents the storyboard to describe a logical process and several scenarios of product usage. Concept Illustration: P5 edits \emph{story} to generate the entire storyboard at once and adjusts \emph{story of single image} to revise images that do not match her expectations. P5 presents the storyboard to illustrate the UI and usage of the product.

\noindent\textbf{P6:} Concept ideation: P6 edits simple sentences in \emph{story} to conclude the ideal product and redesign the product based on the AI-generated results from these comprehensive descriptions. P6 edits \emph{prompts of single image} to iterate each image and finish the logical and rigorous process of product usage. Concept Illustration: Same as above.

\noindent\textbf{P7:} Concept ideation: P7 edits \emph{story} to generate the entire storyboard at once and regenerate the images until they match her expectations. P7 presents the storyboard to illustrate a detailed process of product usage. Concept Illustration: P7 edits \emph{story} to depict each image in the storyboard, and regenerate the images until they match her expectations. P7 presents the storyboard to illustrate the product's UI and usage in detail.

\noindent\textbf{P8:} Concept ideation: P8 edits \emph{story} to depict each image in the storyboard, rather than generating the entire storyboard at once. P8 also regenerates the images until they reflect the product's effectiveness through the emotional shift in users before and after using it. Concept Illustration: P8 edits \emph{story} to depict and revise each image in the storyboard and takes precise control over each image to show the UI and usage of the product.

\noindent\textbf{P9:} Concept ideation: P9 edits \emph{story} to generate the entire storyboard at once and revise \emph{story of single image} to adjust elements in each image that do not match his expectations. P9 takes careful revision over each image to illustrate a detailed story of how they changed after using the product. Concept Illustration: P9 edits \emph{story} to generate the entire storyboard at once and revise \emph{story of single image} to adjust elements in each image that do not match his expectations. P9 also regenerates the images many times to align the storyboard with key elements of the design illustration.

\noindent\textbf{P10:} Concept ideation: P10 edits simple sentences in \emph{story} to present the product usage scenarios and rethink by crafting a narrative for product usage based on AI-generated images. P10 revises her story rather than aligning the images with her original story. Concept Illustration: Same as above.

\noindent\textbf{P11:} Concept ideation: P11 edits \emph{story} to generate the entire storyboard at once and revise \emph{story of single image} to adjust elements in each image that do not match his expectations. P11 takes a precise revision of each image to illustrate a detailed process of product usage. Concept Illustration: P11 edits \emph{story} to generate the entire storyboard at once and revise \emph{prompts of single image} to adjust elements in each image that do not match his expectations. P11 takes a precise revision of each image to illustrate the UI and usage of the product.

\noindent\textbf{P12:} Concept ideation: P12 edits \emph{story} to generate the entire storyboard at once and revise \emph{story of single image} to adjust elements in each image that do not match his expectations. P12 presents the storyboard to describe a logical process and several scenarios of product usage. Concept Illustration: Same as above.

%\end{comment}
%TC:endignore
\end{document}